\documentclass[12pt,fleqn]{article}
\topmargin - 0.8cm

\renewcommand{\theequation}{\arabic{section}.\arabic{equation}}
\renewcommand{\thesection}{\arabic{section}.}

\catcode`@=11 \@addtoreset{equation}{section} \catcode`@=12
\mathchardef\SGamma="7100

\usepackage{amsfonts}
\usepackage{color}
\usepackage{amssymb}
\usepackage{amsmath}

\begin{document}
\title{\vskip-1.7cm \bf  Quantum Effective Action in Spacetimes with
Branes and Boundaries: Diffeomorphism Invariance}
\date{}
\author{A.O.Barvinsky}
\maketitle \hspace{-8mm}{\em Theory Department, Lebedev Physics
Institute, Leninsky Prospect 53, Moscow 119991, Russia}
\begin{abstract}
We construct a gauge-fixing procedure in the path integral for
gravitational models with branes and boundaries. This procedure
incorporates a set of gauge conditions which gauge away effectively
decoupled diffeomorphisms acting in the $(d+1)$-dimensional bulk and
on the $d$-dimensional brane. The corresponding gauge-fixing factor
in the path integral factorizes as a product of the bulk and brane
(surface-theory) factors. This factorization underlies a special
bulk wavefunction representation of the brane effective action. We
develop the semiclassical expansion for this action and explicitly
derive it in the one-loop approximation. The one-loop brane
effective action can be decomposed into the sum of the gauge-fixed
bulk contribution and the contribution of the pseudodifferential
operator of the brane-to-brane propagation of quantum gravitational
perturbations. The gauge dependence of these contributions is
analyzed by the method of Ward identities. By the recently suggested
method of the Neumann-Dirichlet reduction the bulk propagator in the
semiclassical expansion is converted to the Dirichlet boundary
conditions preferable from the calculational viewpoint.
\end{abstract}

\section{Introduction}
In this paper we extend the method of quantum effective action in
brane models \cite{qeastb} to gravitational systems invariant under
general coordinate diffeomorphisms. Specifically, we will be
interested in the peculiarities of the gauge-fixing procedure caused
by the presence of branes/boundaries. It is well known that branes
break the full diffeomorphism invariance and give rise to the
dynamical brane bending modes \cite{GarrigaTanaka} which produce
ghost instabilities \cite{ghosts,ghoststachions} and generate a low
strong-coupling scale \cite{scale}. This leads to difficulties in
the construction of long-distance modifications of gravity theory
that could underly the dark energy phenomenon both at the classical
and quantum level \cite{NicolisRattazzi}. Quantum effects in
gravitational and, in particular, cosmological brane models might
also be important for the mechanism of the cosmological acceleration
that can be facilitated by the 4D conformal anomaly in the presence
of extra dimensions \cite{slih}. So here we want to study the
gauge-fixing aspects of such problems in the quantum domain.

Peculiarities of the gauge-fixing procedure in brane models follow
from the structure of their action
    \begin{equation}
    S\,[\,G_{AB}(X)\,]=
    S^{\rm B}[\,G_{AB}(X)\,]
    +S^{\rm b}[\,g_{\alpha\beta}(x)\,].          \label{1.1}
    \end{equation}
It contains the bulk and brane parts as functionals of the bulk
metric $G_{AB}(X)$ and the induced metric on the brane
$g_{\alpha\beta}(x)$. The bulk $(d+1)$-dimensional and the brane
$d$-dimensional coordinates are labeled respectively by $X=X^A$,
$A=0,1,...d$, and $x=x^\alpha$, $\alpha=0,1,...d-1$, the brane is
embedded into the bulk by means of some embedding function $e(x)$,
    \begin{eqnarray}
    X^A=e^A(x),      \label{3.6}
    \end{eqnarray}
and the induced metric reads in terms of the bulk metric as
    \begin{eqnarray}
    g_{\alpha\beta}(x)=
    \partial_\alpha e^A(x)\,G_{AB}(e(x))\,
    \partial_\beta e^B(x).                      \label{3.8}
    \end{eqnarray}

Generically, the bulk part in (\ref{1.1}) is the Einstein action
with the cosmological term
     \begin{eqnarray}
     &&S^{\rm B}[\,G_{AB}(X)\,]=\frac1{16\pi G}\left\{
     \int_{\rm\bf B} d^{d+1}X\,G^{1/2}
     \big(R(G)-2\Lambda \big)
     -2\int_{\rm\bf b} d^dx\,
     g^{1/2}\,K\right\}           \label{action}
     \end{eqnarray}
which includes the integral over the bulk {\bf B} and the
Gibbons-Hawking integral over the brane {\bf b}. The latter involves
the trace of the brane extrinsic curvature
$K=g^{\alpha\beta}K_{\alpha\beta}$ and is necessary for the
consistency of the variational procedure for this
action.\footnote{The extrinsic curvature is defined as a projection
onto the brane of the tensor $K_{AB}\equiv\nabla_A n_B$ with the
{\em inward} pointing vector $n_B$ normal to the brane.}

The brane part of (\ref{action}) is given by a covariant
$d$-dimensional integral over {\bf b}. Depending on the model it
contains the brane tension term (in the Randall-Sundrum model
\cite{RS}), the brane Einstein term (like in the
Dvali-Gabadadze-Porrati model \cite{DGP}), the combination thereof
\cite{Tanaka} or other covariant structures in brane and bulk
curvatures \cite{other}. Both bulk and brane parts of the full
action may also contain matter fields which for brevity we do not
consider here. Without loss of generality they can be included into
the sets of bulk and brane metric coefficients. Also, it is worth
noting that Eqs.(\ref{1.1})-(\ref{action}) equally well describe the
system in the bulk domain {\bf B} with the boundary ${\rm\bf
b}=\partial{\rm\bf B}$ and the brane system with the bulk satisfying
the $Z_2$-symmetry with respect to the brane {\bf b}. This formalism
also applies in context of Euclidean quantum gravity with a closed
compact boundary {\bf b}.

The bulk and brane parts of the action (\ref{1.1}) are invariant
with respect to diffeomorphisms generated by the bulk $\Xi^A(X)$ and
brane $\xi^\alpha(x)$ vector fields. The corresponding
transformations have the form
    \begin{eqnarray}
    &&\delta^\Xi G_{AB}(X)=
    \nabla_A\Xi_B(X)+\nabla_B\Xi_A(X),                 \label{1.3}\\
    &&\delta^\xi g_{\alpha\beta}(x)=
    D_\alpha\xi_\beta(x)+D_\beta\xi_\alpha(x),          \label{1.4}
    \end{eqnarray}
where $\nabla_A$ denotes the covariant derivative in the bulk and
$D_\alpha$ is a covariant derivative on the brane. The bulk
diffeomorphism (\ref{1.3}) preserves the bulk action only when the
projection of $\Xi^A(X)$ normal to the brane, $\Xi^\perp(X)$, is
vanishing. In contrast, the tangential projection of $\Xi^A(X)$,
$\Xi^\alpha(X)$, on the brane can be arbitrary. Because the bulk and
induced brane metrics are not independent,  this projection
generates the brane diffeomorphism and coincides with the vector
field $\xi^\alpha(x)$. Thus, the boundary conditions for the above
gauge transformations read
    \begin{eqnarray}
    \Xi^\alpha(X) \big|=\xi^\alpha(x),\,\,\,
    \Xi^\perp(X) \big|=0.                        \label{1.5}
    \end{eqnarray}
Here and in what follows the vertical bar indicates that the
function in the bulk, $\phi(X)$, is restricted to the brane and
labeled by the corresponding low case letter $x$,
$\phi(X)|=\phi(e(x))$, with the aid of the embedding function
(\ref{3.6}).

The construction of the quantum effective action for any brane model
should incorporate the gauge-fixing procedure for the diffeomorphism
symmetry (\ref{1.3})-(\ref{1.5}). Schematically, this corresponds to
introducing the Feynman-DeWitt-Faddeev-Popov gauge-fixing factor in
the path integral \cite{DW}
    \begin{equation}
    e^{\,\textstyle i\SGamma}=\int
    DG_{AB}(X)\,\exp\big(\,iS\,[\;G_{AB}(X)\,]\big)
    \times(\,\mbox{gauge-fixing}\,).                  \label{1.2}
    \end{equation}
Here integration runs over the metric coefficients in the bulk and
on the brane because the metric on the brane is not fixed and is
subject to quantum fluctuations. The gauge-fixing factor should
gauge away local diffeomorphisms by imposing a certain set of gauge
conditions and introducing necessary ghost factors maintaining
unitarity and gauge independence of the effective action.

Curiously, despite a well-known form of the gauge algebra
(\ref{1.3})-(\ref{1.5}), no exhaustive formulation of gauge-fixing
procedure for (\ref{1.2}) is known in current literature. In
particular, the ghost factors require the knowledge of boundary
conditions for the corresponding Faddeev-Popov operators acting on
the space of gauge parameters $\Xi^A(X)$. In view of (\ref{1.5})
their tangential components $\Xi^\alpha(X)|$ are arbitrary and
should be integrated over, and it would seem that this integration
gives rise to the Neumann boundary conditions for $\Xi^\alpha(X)|$.
However, this naive prescription turns out to be incorrect. As we
show below, the derivation of the boundary conditions for ghosts is
less straightforward and much deeper relies on the details of
gauging away the transformations (\ref{1.3})-(\ref{1.5}).

In its turn, the preferable choice of gauge conditions in brane
models is constrained by additional requirements. One requirement
follows from the fact that within the brane concept the fields in
the bulk are usually integrated out, so that the physics of the
system is effectively probed only by the variables living on the
brane.\footnote{This means that in the classical action of the path
integral (\ref{1.2}) one must add sources conjugated only to the
variables located at the brane, that is to the components of the
induced metric $g_{\alpha\beta}(x)$. The Legendre transform with
respect to these sources then leads to the effective action as a
functional of the mean {\em induced} metric, $\varGamma[\,\langle
g_{\alpha\beta}(x)\rangle\,]$. As is well known, in the one-loop
approximation this field is called the background, and the whole
procedure reduces to the subtraction from $S\,[\;G_{AB}(X)\,]$ in
(\ref{1.2}) the term linear in the deviation of the integration
variable from this background -- the stationary point of the path
integral.} Thus, a very important aspect of the effective action in
brane theory is that it should be a functional of the induced metric
$g_{\alpha\beta}(x)$, rather then the full set of metric
coefficients $G_{AB}(X)$. Therefore, the preferable gauge conditions
should be chosen in such a way that the calculational technique for
$\varGamma[\, g_{\alpha\beta}(x)\,]$ is manifestly covariant with
respect to brane diffeomorphisms (\ref{1.4}) {\em separately} from
the bulk ones (\ref{1.3}).

The effective decoupling of brane diffeomorphisms from the bulk
diffeomorphisms can be attained by imposing two sets of gauge
conditions --- brane $d$-dimensional gauges $\chi^\mu(x)=0$ and the
bulk $(d+1)$-dimensional gauges $H^A(X)=0$ which gauge away their
corresponding diffeomorphisms. Moreover, they can preserve the
manifest covariance of the calculational scheme if they are chosen
in the DeWitt background-covariant form \cite{DW}. Under the
splitting of the brane metric into the background (mean field) and
perturbation parts, $g_{\alpha\beta}\to
g_{\alpha\beta}+h_{\alpha\beta}$, this brane gauge reads as
    \begin{eqnarray}
    \chi^\mu=D^\nu h_{\mu\nu}-\frac12 D^\mu h,
    \,\,\,h\equiv g^{\mu\nu}\,h_{\mu\nu},          \label{4.9}
    \end{eqnarray}
where the covariant derivatives $D_\mu$ are defined with respect to
the background metric $g_{\mu\nu}$ which also raises the indices of
$h_{\mu\nu}$ and $D_\mu$. The corresponding Faddeev-Popov operator
which we denote by $\mbox{\boldmath$J$}^\mu_\nu(D)$ acts on the
vector space of $\xi^\nu(x)$. In the lowest order in $h_{\mu\nu}$ it
reads as a covariant $d$-dimensional d'Alembertian modified by the
Ricci-curvature potential term
    \begin{eqnarray}
    \mbox{\boldmath$J$}^\mu_\nu(D)=\Box^{(d)}\delta^\mu_\nu
    +R^\mu_\nu,\,\,\,\,
    \Box^{(d)}=g^{\mu\nu}D_\mu D_\nu.              \label{4.10}
    \end{eqnarray}

A similar DeWitt bulk gauge condition arises under the splitting of
the full bulk metric, $G_{AB}\to G_{AB}+H_{AB}$, and reads
    \begin{eqnarray}
    H^A=\nabla^B H_{BC}-\frac12 \nabla^A H,
    \,\,\,H\equiv G^{AB}\,H_{AB},          \label{4.6}
    \end{eqnarray}
where $\nabla_A$ is also defined with respect to the background
metric $G_{AB}$. The corresponding Faddeev-Popov operator, denoted
by $Q^A_{B}(\nabla)$ and acting on the space of $\Xi^B(X)$, is given
by
    \begin{eqnarray}
    Q^A_{B}(\nabla)=\Box^{(d+1)}\delta^A_B+R^A_B,\,\,\,\,
    \Box^{(d+1)}=\nabla^A\nabla_A.          \label{4.7}
    \end{eqnarray}
This operator is of the second order in derivatives and, as we show
below, it should be supplied by the Dirichlet boundary conditions
for {\em all} components of $\Xi^A(X)$. This guarantees the
decoupling of the bulk and brane diffeomorphisms even despite their
entanglement via the relation (\ref{1.5}).

Another requirement to the background field formalism for brane
theories is the reduction of their Feynman rules to those of the
Green's functions with Dirichlet boundary conditions. As discussed
in \cite{qeastb} such boundary conditions are calculationally much
simpler than the original brane matching conditions of the
generalized Neumann type. This reduction is possible in view of the
duality relations between the Dirichlet and Neumann boundary value
problems suggested in \cite{duality}, and as we show below this
reduction also works for gauge-fixed framework. In particular, the
decomposition of the quantum effective action in the sum of the bulk
and brane-to-brane contributions obtained in \cite{qeastb}, which
incorporates this Neumann-Dirichlet reduction, works both in the
gauge-field and ghost sectors of (\ref{1.2}).

Below we develop the gauge-fixing procedure of the above type. We
derive the gauge-fixing factor in Eq.(\ref{1.2}), show its
bulk-brane factorization property underlying a special bulk
wavefunction representation of the brane effective action
\cite{gospel} and build the semiclassical expansion for this action
with its Neumann-Dirichlet reduction.

The paper is organized as follows. In Sect.2 we list the main
results of this paper. Sect.3 presents the spacetime foliation
associated with its brane/boundary and the corresponding canonical
formalism which facilitates the derivation of relevant boundary
value problems. In Sects. 4 and 5 we derive the gauge-fixing factor
of (\ref{1.2}) and the associated bulk wavefunction representation
of the brane effective action. In Sect. 6 we derive the Feynman
diagrammatic technique for this action and explicitly present it in
the one-loop approximation as a sum of special bulk and brane
contributions. They are based on a mixed Dirichlet-Neumann boundary
value problem for the graviton propagator. Sect.7 presents the
method of Ward identities which allow one to demonstrate for both of
these contributions their manifest gauge independence. In Sect. 8
the diagrammatic technique is completely reduced to the propagator
with strictly Dirichlet boundary conditions. The concluding section
summarizes the obtained results and discusses their possible
applications. Three appendices describe the properties of the local
measure in the path integral for brane models, the Gaussian
integration over the functional space with mixed Dirichlet-Neumann
boundary conditions and the variational problem for the Green's
function subject to such boundary conditions.

\section{Main results}
The main results of this paper include the selection of bulk and
brane gauge conditions, their corresponding gauge-fixing factor in
the path integral (\ref{1.2}), the one-loop expression for the brane
effective action and its Neumann-Dirichlet reduction. In this
section we briefly list these results that will be derived in the
following sections.

\subsection{Bulk and brane gauge conditions}
In view of (\ref{1.5}) a generic diffeomorphism in the bulk can be
represented as a composition of two transformations. One
transformation is uniquely fixed by a given vector field
$\xi^\mu(x)$ on the brane by continuing its $d$-dimensional
diffeomorphism from the brane to the spacetime bulk. Another
transformation is a generic $(d+1)$-dimensional diffeomorphism which
reduces to the identity at the boundary, $\Xi^A(X)\,|=0$.
Correspondingly, the gauge-fixing procedure can be split into two
stages. At the first stage special bulk gauge conditions gauge away
this $(d+1)$-dimensional diffeomorphisms vanishing at the brane. At
the second stage brane gauge conditions completely fix the residual
$d$-dimensional diffeomorphisms at the boundary. This strategy
allows one to decouple the boundary diffeomorphisms from the bulk
ones and render the whole formalism manifestly covariant with
respect to the both types of diffeomorphisms.

For this purpose the bulk gauge conditions should be special in the
sense that they should not overconstrain the relevant
$(d+1)$-dimensional transformations and leave the boundary values
$\Xi^\mu(X)\,|=\xi^\mu(x)$ arbitrary. This means that $\xi^\mu(x)$
play the role of boundary conditions for zero modes of the relevant
Faddeev-Popov operator --- the transformations which leave the bulk
gauge conditions invariant. If we denote these $(d+1)$-dimensional
gauge conditions by
    \begin{eqnarray}
    H^A(X)=H^A\big(G_{CD}(X),
    \partial_B G_{CD}(X)\big)    \label{2.1}
    \end{eqnarray}
then their gauge transformation serves as the definition of the bulk
Faddeev-Popov operator $Q^A_B(\nabla)$
    \begin{eqnarray}
    \delta^\Xi H^A(X)
    =Q^A_B(\nabla)\,\Xi^B(X).              \label{2.3}
    \end{eqnarray}
To fix the bulk diffeomorphisms this operator should be invertible
under the Dirichlet boundary conditions $\Xi^A(X)\,|=0$, but it must
have nontrivial zero modes $\Xi^B_0(X)$,
$Q^A_B(\nabla)\,\Xi^B_0(X)=0$, subject to inhomogeneous boundary
conditions (\ref{1.5}) ``enumerated" by all possible brane vector
fields $\xi^\mu(x)$. This is possible when the Faddeev-Popov
operator in (\ref{2.3}) is of the second order in derivatives
transversal to the spacetime boundary. The gauge conditions which
generate the Faddeev-Popov operators with such properties can be
background-covariant from the bulk $(d+1)$-dimensional viewpoint
and, in particular, given by the DeWitt gauge in the bulk
(\ref{4.6}).

The residual gauge transformations with the parameters $\xi^\mu$ are
gauged away by imposing the brane gauge conditions $\chi^\mu(x)=0$.
In order to decouple the boundary diffeomorphisms from the bulk
ones, one must impose these gauge conditions only upon the brane
metric
    \begin{eqnarray}
    \chi^\mu(x)=
    \chi^\mu\big(g_{\alpha\beta}(x),
    \partial_\mu g_{\alpha\beta}(x)\big).  \label{2.2}
    \end{eqnarray}
Similarly to (\ref{2.3}) they generate the brane Faddeev-Popov
operator $\mbox{\boldmath$J$}^\mu_\nu(D)$,
    \begin{eqnarray}
    \delta^\xi \chi^\mu(x)
    =\mbox{\boldmath$J$}^\mu_\nu(D)\,
    \xi^\nu(x),                               \label{2.4}
    \end{eqnarray}
which is now determined entirely in terms of quantities induced on
the brane. A particular example of such gauge conditions which are
background-covariant from the $d$-dimensional viewpoint is given by
the DeWitt gauge conditions (\ref{4.9}) with the Faddeev-Popov
operator (\ref{4.10}) which is nondegenerate under appropriate
boundary conditions at the infinity of the brane (or in view of the
closed compact nature of the boundary in Euclidean context when
$\Box^{(d)}$ becomes a Laplacian).

\subsection{The gauge-fixing factor}
For the gauge-fixing procedure (\ref{2.1})-(\ref{2.4}) the
gauge-fixing factor in (\ref{1.2}) factorizes into the product of
the corresponding bulk and brane factors \cite{gospel}
    \begin{equation}
    (\,\mbox{gauge-fixing}\,)=
    \delta\,[\,H\,]\;{\rm Det}_{\rm D}Q\times\,
    \delta(\chi)\,
    \det\mbox{\boldmath$J$}.                  \label{2.5}
    \end{equation}
Here $\delta\,[\,H\,]$ and $\delta(\chi)$ denote respectively the
$(d+1)$-dimensional and $d$-dimensional functional delta-functions,
    \begin{eqnarray}
    &&\delta [\,H\,]\equiv\prod\limits_{X\in{\rm\bf B}}\,
    \prod\limits_A\,
    \delta\big(H^A(X)\big),                    \label{2.6}\\
    &&\delta(\chi)
    \equiv\prod\limits_{x\in{\rm\bf b}}\,
    \prod\limits_\mu\,
    \delta\big(\chi^\mu(x)\big),                  \label{2.7}
    \end{eqnarray}
and ${\rm Det}_{\rm D}Q$ and $\det\mbox{\boldmath$J$}$ are the
corresponding ghost functional determinants of the bulk and brane
Faddeev-Popov operators defined by (\ref{2.3}) and (\ref{2.4}). To
distinguish between the functional dimensionalities of these
determinants we denote the determinant of the $(d+1)$-dimensional
theory by ${\rm Det}\equiv{\rm Det}^{(d+1)}$ and that of the
$d$-dimensional theory by ${\rm det}\equiv{\rm Det}^{(d)}$.

The subscript in ${\rm Det}_{\rm D}Q$ indicates that the functional
determinant of the second-order differential operator
$Q^A_{B}(\nabla)$ is taken subject to the {\em Dirichlet} boundary
conditions on the brane. The variational definition of such a
determinant in terms of the Dirichlet Green's function of
$Q^A_{B}(\nabla)$ is given below, see Eq.(\ref{4.13a}). In contrast,
boundary conditions for $\mbox{\boldmath$J$}(D)$ in
$\det\mbox{\boldmath$J$}$ are not important for us, because the
brane boundary is either absent, as in Euclidean context with a
closed compact boundary of a bulk domain (the boundary of the
boundary), or lies at infinity where the boundary conditions are
trivial.

One can use the t'Hooft's method of transition from the degenerate
(delta-function type) gauges to the gauge-breaking terms both in the
bulk sector
    \begin{eqnarray}
    &&\delta\left[\,H\,\right]\,\Rightarrow
    \left(\,{\rm Det}\,c_{AB}\,\right)^{1/2}\,
    \exp\Big(i S^{\rm B}_{\rm gb}[\,G\,]\Big),  \label{2.8}\\
    &&S^{\rm B}_{\rm gb}[\,G(X)\,]=-\frac 12\,
    \int_{\bf B} dX\, H^A(X)\,
    c_{AB}\,H^B(X),                             \label{2.9}
    \end{eqnarray}
and brane sector
    \begin{eqnarray}
    &&\delta\left(\,\chi\right)\,\Rightarrow
    \big(\,{\rm det}\,c_{\mu\nu}\,\big)^{1/2}\,
    \exp\Big( i S^{\rm b}_{\rm gb}[\,g\,]\Big),  \label{2.10}\\
    &&S^{\rm b}_{\rm gb}[\,g(x)\,]
    =-\frac 12\,\int_{\bf b}\,dx\,\chi^\mu(x)\,
    c_{\mu\nu}\,\chi^\nu(x).                    \label{2.10a}
    \end{eqnarray}
Here $c_{AB}$ and $c_{\mu\nu}$ are respectively the bulk and brane
gauge-fixing matrices. We use for them the same notation differing
only by the type of indices ($AB$ vs $\mu\nu$), which should not
lead to a confusion in what follows.

The factorization of the gauge-fixing factor (\ref{2.5}) underlies a
special bulk wavefunction representation of the brane effective
action, see Eqs. (\ref{5.2})-(\ref{5.3}) below.

\subsection{One-loop brane effective action}
The semiclassical expansion for the brane effective action within
the gauge-fixing procedure of the above type has the form
    \begin{eqnarray}
    &&\SGamma[\,g_{\alpha\beta}(x)\,]=
    S^{\rm B}\big[\,G^0_{AB}[\,g(x)\,]\,\big]+
    S^{\rm b}[\,g_{\alpha\beta}(x)\,]\nonumber\\
    &&\qquad\qquad\qquad\qquad\qquad\quad
    +\hbar\,\SGamma_{\rm 1-loop}
    [\,g_{\alpha\beta}(x)\,]
    +O(\,\hbar^2\,).                          \label{2.11}
    \end{eqnarray}
Its tree-level part (the first line of this equation) follows from
the classical action (\ref{1.1}) calculated on the solution of
classical equations of motion, $G^0_{AB}[\,g(x)\,]$. This solution
satisfies the bulk gauge (\ref{2.1}) and is subject to the boundary
condition on the brane --- the induced $d$-dimensional metric
$g_{\alpha\beta}(x)$,
    \begin{eqnarray}
    &&\left.\frac{\delta S^{\rm B}[\,G\,]}
    {\delta G_{AB}(X)}\,\right|_{\,\,G=G^0}=0,   \label{2.12}\\
    &&H^A\big(G^0(X),\partial G^0(X)\big)=0,    \label{2.12a}\\
    &&G^0_{\alpha\beta}(X)\big|_{\,\rm b}
    =g_{\alpha\beta}(x).                          \label{2.12b}
    \end{eqnarray}

The one-loop part of the action is built in terms of the inverse
propagator of the theory in the bulk, which is non-degenerate due to
the contribution of the bulk gauge-breaking term
    \begin{eqnarray}
    F=F^{AB,\,CD}(\nabla)\,\delta(X,X')=
    \frac{\delta^2
    (\,S^{\rm B}+S^{\rm B}_{\rm gb}\,)}
    {\delta G_{AB}(X)\,\delta G_{CD}(X')}.    \label{2.13}
    \end{eqnarray}
The Green's function of this operator, which determines the bulk
propagator,
    \begin{eqnarray}
    G_{\rm DN}(X,X')=G_{\,\,AB,\,CD}^{\rm DN}(X,X'),  \label{2.14}
    \end{eqnarray}
satisfies the following problem with the set of mixed boundary
conditions of the generalized Dirichlet-Neumann type
    \begin{eqnarray}
    &&F(\nabla)\,G_{\rm DN}(X,X')=\delta(X,X'),   \label{2.15}\\
    &&G_{\,\,\alpha\beta,\,CD}^{\rm DN}(X,X')
    \big|_{\,X}=0,                               \label{2.15a}\\
    &&\stackrel{\rightarrow}
    {{\vphantom W}H}\!(\nabla)\,
    G_{\rm DN}(X,X')\big|_{\,X}=0.               \label{2.15b}
    \end{eqnarray}
Here (\ref{2.15a}) means that the induced metric components of the
Green's function ($\alpha\beta$-components among $AB$-components)
vanish on the brane, whereas the rest of the boundary conditions
imply the vanishing of the linearized gauge conditions. The
differential operator of linearized gauge conditions, acting upon
the Green's function in (\ref{2.15b}),
$H^{E,\,AB}(\nabla)G_{\,\,AB,\,CD}^{\rm DN}(X,X')$, is defined by
    \begin{eqnarray}
    \stackrel{\rightarrow}{{\vphantom W}H}\!(\nabla)\,\delta(X,X')
    \equiv\;\stackrel{\rightarrow}{{\vphantom W}H}\!
    {\vphantom H}^{E,\,AB}(\nabla)\,\delta(X,X')=
    \frac{\delta H^E(X)}{\delta G_{AB}(X')}.     \label{2.16}
    \end{eqnarray}

In terms of these quantities the one-loop effective action reads as
the sum of the bulk and brane effective actions both including their
relevant ghost contributions (corresponding to the factorization of
the gauge-fixing factor (\ref{2.5}))
    \begin{eqnarray}
    &&i\SGamma_{\rm 1-loop}[\,g_{\alpha\beta}(x)\,]=
    -\frac12\,{\rm Tr}_{\rm DN}\ln F
    +{\rm Tr}_{\rm D}\ln Q
    \,
    \Big|_{\;G\,=\;G_0(\,g)}\nonumber\\
    &&\qquad\qquad\qquad\qquad\qquad\qquad
    -\frac12\,{\rm tr}\,\ln \mbox{\boldmath$F$}^{\rm DN}
    +{\rm tr}\,\ln \mbox{\boldmath$J$}.              \label{2.17}
    \end{eqnarray}
Here ${\rm Tr}_{\rm D}$ and ${\rm Tr}_{\rm DN}$ denote the
functional traces of the bulk theory subject to the Dirichlet and
mixed Dirichlet-Neumann (cf. (\ref{2.15})-(\ref{2.15b})) boundary
conditions, while ${\rm tr}$ is a functional trace in the boundary
$d$-dimensional theory. $\mbox{\boldmath$F$}^{\rm DN}$ is the
brane-to-brane operator
    \begin{eqnarray}
    &&\mbox{\boldmath$F$}^{\rm DN}\equiv
    \mbox{\boldmath$F$}^{\alpha\beta,\,
    \gamma\delta}_{\rm DN}(x,x')\nonumber\\
    &&\qquad\qquad=
    -\big( \stackrel{\rightarrow}{{\vphantom W}W}\!G_{\rm DN}\!
    \stackrel{\leftarrow}{{\vphantom W}W}\!||\,\big)
    ^{\alpha\beta,\,\gamma\delta}\,(x,x')
    +\kappa^{\alpha\beta,\,
    \gamma\delta}(D)\,\delta(x,x').              \label{2.19}
    \end{eqnarray}

This is a gauge theory generalization of the brane operator
introduced in \cite{qeastb}. The first term here implies that the
kernel of the Dirichlet-Neumann Green's function is being acted by
the operators $W(\nabla)$ upon both arguments in the directions
indicated by arrows. The double vertical bar indicates that both
points of the kernel are restricted to the brane and labeled by the
corresponding low case letters. That is, if the embedding of the
boundary/brane in the bulk is denoted by $X=e(x)$, then this
explicitly means:
    \begin{eqnarray}
    \stackrel{\rightarrow}
    {{\vphantom W}W}\!G_{\rm DN}\!
    \stackrel{\leftarrow}
    {{\vphantom W}W}\!||\,(x,x')\equiv\;
    \stackrel{\rightarrow}
    {{\vphantom W}W}\!(\nabla)\,G_{\rm DN}(X,X')\!
    \stackrel{\leftarrow}
    {{\vphantom W}W}\!(\nabla')
    \,\Big|_{\,X=e(x),\,X'=e(x')}.             \label{2.20}
    \end{eqnarray}
The first order differential operator $W(\nabla)$ is the Wronskian
operator
    \begin{eqnarray}
    \stackrel{\rightarrow}{{\vphantom W}W}\!\!(\nabla)
    =\;\stackrel{\rightarrow}{{\vphantom W}W}\!\!
    {\vphantom W}^{AB,\,CD}(\nabla),            \label{2.21}
    \end{eqnarray}
which is determined for the symmetric second-order differential
operator (\ref{2.13}) by the following Wronskian relation valid for
two arbitrary test functions $\phi_{1,2}(X)$
    \begin{eqnarray}
    \int\limits_{\rm B} dX
    \left(\phi_1\!
    \stackrel{\rightarrow}{{\vphantom W}F}
    \!\!(\nabla)\phi_2-
    \phi_1\!
    \stackrel{\leftarrow}{{\vphantom W}F}
    \!\!(\nabla)\,\phi_2\right)=
    -\int\limits_{\rm \partial B} dx
    \left(\phi_1\!
    \stackrel{\rightarrow}{{\vphantom W}W}
    \!\!(\nabla)\phi_2-
    \phi_1\!
    \stackrel{\leftarrow}{{\vphantom W}W}
    \!\!(\nabla)\,\phi_2\right).               \label{2.22}
    \end{eqnarray}

With these definitions the first term of (\ref{2.19}) is given by
the $(\alpha\beta,\,\gamma\delta)$-block of the matrix $(
\stackrel{\rightarrow}{{\vphantom W}W}\!G_{\rm DN}\!
\stackrel{\leftarrow}{{\vphantom W}W}\!||\,)^{AB,\,CD}$. The second
term of (\ref{2.19}) is a contribution of the brane $d$-dimensional
part of the classical action (\ref{1.1}) and the brane
gauge-breaking term (\ref{2.10a})
    \begin{eqnarray}
    \kappa^{\alpha\beta,\,\gamma\delta}(D)\,\delta(x,x')=
    \frac{\delta^2 (\,S^{\rm b}
    +S^{\rm b}_{\rm gb}\,)}
    {\delta g_{\alpha\beta}(x)
    \,\delta g_{\gamma\delta}(x')}.             \label{2.23}
    \end{eqnarray}
The gauge-breaking term here makes the whole operator (\ref{2.19})
nondegenerate.

\subsection{Neumann-Dirichlet reduction}
The equation (\ref{2.17}) generalizes the technique of \cite{qeastb}
to brane gravitational models invariant with respect to local
diffeomorphisms. The main goal of \cite{qeastb} was a complete
reduction of the calculational technique from Neumann-type boundary
conditions to much simpler Dirichlet ones. However, in the algorithm
(\ref{2.17}) this goal is not yet achieved, because of the
complexity of boundary conditions for $G_{\rm DN}(X,X')$ in
(\ref{2.15})-(\ref{2.15b}). Such a reduction can be done, and in the
one-loop approximation it is given by the expression alternative to
(\ref{2.17})
    \begin{eqnarray}
    &&i\SGamma_{\rm 1-loop}[\,g_{\alpha\beta}(x)\,]=
    -\frac12\,{\rm Tr}_{\rm D}\ln F
    +{\rm Tr}_{\rm D}\ln Q
    \nonumber\\
    &&\qquad\qquad\qquad\qquad\qquad\qquad\qquad\quad
    -\frac12\,{\rm tr}\,
    \ln \mbox{\boldmath$F$}^{\rm D}
    +{\rm tr}\,\ln\mbox{\boldmath$J$}.              \label{8.19}
    \end{eqnarray}
Here the functional traces of all bulk operators are calculated
subject to Dirichlet conditions, and the brane-to-brane operator
$\mbox{\boldmath$F$}^{\rm D}$ is given by the expression similar to
(\ref{2.19}) but with the Dirichlet Green's function $G_{\rm D}$
replacing $G_{\rm DN}$
    \begin{eqnarray}
    \mbox{\boldmath$F$}_{\rm D}^{\alpha\beta,\,
    \gamma\delta}(x,x')=
    -\big( \stackrel{\rightarrow}{{\vphantom W}W}\!G_{\rm D}\!
    \stackrel{\leftarrow}{{\vphantom W}W}\!||\,\big)
    ^{\alpha\beta,\,\gamma\delta}\,(x,x')
    +\kappa^{\alpha\beta,\,
    \gamma\delta}(D)\,\delta(x,x').              \label{2.19a}
    \end{eqnarray}

\section{Canonical formalism in the brane foliation of spacetime}
The derivation of the above results is much easier in terms of the
DeWitt condensed notations \cite{DW}. In these notations the bulk
metric is labeled by the condensed index $a=(AB,X)$ including both
tensor labels and the {\em bulk coordinates} $X$
    \begin{eqnarray}
    G^a=G_{AB}(X),\,\,\,a=(AB,X).     \label{3.1}
    \end{eqnarray}
If we apply the same convention to the definition of the vector
field $\Xi^A=\Xi^A(X)$, $A=(A,X)$, then the gauge invariance of the
bulk action can be written down in terms of the generators of
diffeomorphisms $R^a_{\;A}$
    \begin{eqnarray}
    &&\delta^\Xi G^a=R^a_{\,\;A} \,\Xi^A,      \label{3.2}\\
    &&R^{\,\;\;a}_A\,\frac{\delta S^{\rm B}}
    {\delta G^a}=0,                          \label{3.3}
    \end{eqnarray}
where the contraction of condensed indices implies also integration
over the bulk coordinates $X$. In order to display the differential
structure of these generators for various metric components
    \begin{eqnarray}
    R^a_{\,C}=\,\,
    \stackrel{\rightarrow}{{\vphantom W}R}{\!\vphantom R}_{AB,\,C}
    (\nabla)\,\delta(X,X'),\,\,\,
    \stackrel{\rightarrow}{{\vphantom W}R}
    {\!\vphantom R}_{AB,\,C}(\nabla)
    =2G_{(AC}\nabla_{B)},                     \label{3.4}
    \end{eqnarray}
we will need the canonical formalism of the bulk gravitational
action, associated with the brane-type foliation of the full
spacetime. In this formalism the role of time is played by the
extra-dimensional coordinate $y$. In view of the spacelike nature of
$y$ this variable has nothing to do with a real dynamical evolution,
but the local differential properties of the diffeomorphism
transformations are very similar to those of the usual canonical
formalism in the physical time $t=X^0$.

So we assume that the bulk is foliated by the surfaces of constant
$y$, whose embedding is determined by the embedding functions,
    $X^A=e^A(x^\alpha,y)$,
including the embedding of the brane (\ref{3.6}),
$e^A(x^\alpha)\equiv e^A(x^\alpha,0)$. Such a foliation determines
the vielbein of vectors $e^A_\alpha$ and $n^A$ respectively
tangential and normal to the slices,
    $e^A_\alpha\equiv\partial_\alpha e^A,\,\,\,\,
    n_A\,e^A_\alpha=0,\,\,G_{AB}\,n^A n^B=1$.
It also gives rise to the induced metric on the brane and other
surfaces of constant $y$ -- the generalization of Eq.(\ref{3.6}),
    $g_{\alpha\beta}(x,y)=
    e^A_\alpha(x,y)\,G_{AB}(e(x,y))\,e^B_\beta(x,y)$.

This foliation generates the lapse and shift functions defined as
normal and tangential projections of the local ``velocity" vector
$\partial_y e^A(x,y)\equiv\dot e^A$ with which the bulk slice
evolves in ``time" $y$
    \begin{eqnarray}
    N^\perp=n_A\,\partial_y e^A(x,y),\,\,\,
    N^\alpha=g^{\alpha\beta}\,N_\beta,\,\,\,
    N_\beta=e^B_\beta\,G_{BA}\,
    \partial_y e^A(x,y).                            \label{3.9}
    \end{eqnarray}
In fact these functions are equivalent to $G_{Ay}$-components of the
full metric, so that $G_{AB}(X)$ can be parameterized in terms of
$g_{\alpha\beta}$ and $N^A=(N^\alpha, N^\perp)$. Then the bulk
Einstein action (\ref{action}) has a well-known ADM form in terms of
the extrinsic curvature $K_{\alpha\beta}$ of constant $y$ slices and
their scalar curvature $R^{(d)}(g)$
    \begin{eqnarray}
    &&S^{\rm B}[\,G\,]=\frac1{16\pi G}\int\limits_{y\geq 0}
    dy\,\int dx\,g^{1/2}\,
    N^\perp\,\big(K^2-K_{\alpha\beta}^2
    +R^{(d)}(g)-2\Lambda\big),                              \label{3.10}\\
    &&K_{\alpha\beta}=\frac1{2N^\perp}
    \big(\,\partial_y g_{\alpha\beta}
    -D_\alpha N_\beta-D_\beta
    N_\alpha\big).
    \end{eqnarray}
It is important that only the "velocities" of $g_{\alpha\beta}$,
    \begin{eqnarray}
    \partial_y g_{\alpha\beta}
    \equiv\dot g_{\alpha\beta},  \label{3.11}
    \end{eqnarray}
enter the Lagrangian, while the lapse and shift functions $N^A$ are
not dynamical and serve as Lagrange multipliers in the $y$-time
canonical formalism of the action (\ref{3.10}). In spacetime
condensed notations of (\ref{3.1}) we will denote the induced metric
of $y$-slices by $g^i$ and this decomposition will look like
    \begin{eqnarray}
    &&G^a=(g^i,\,N^A),\,\,\,a=(i,A),\,\,\,
    i=(\alpha\beta,X),\,\,\,
    A=(\alpha,\perp,\;X),              \nonumber\\
    &&g^i=g_{\alpha\beta}(X),\,\,\,\,
    N^A=\big(N^\alpha(X),\,N^\perp(X)\big).     \label{3.12}
    \end{eqnarray}

In what follows we will also need {\em canonical} condensed
notations in which the indices $(a,i,A)$ include together with
discrete labels only the {\em brane coordinates} $x$, and the
contraction of these indices implies the integration only over $x$.
    \begin{eqnarray}
    &&G^a(y)=\big(g^i(y),\,N^A(y)\big),
    \,\,\,i=(\alpha\beta,x),\,\,\,
    A=(\alpha,\perp,\;x),                     \nonumber\\
    &&g^i(y)=g_{\alpha\beta}(x,y),\,\,\,\,\,
    N^A(y)=(N^\alpha(x,y),\,N^\perp(x,y))     \label{3.13}
    \end{eqnarray}
In these notations the generators of Eq.(\ref{3.2}), $R^a_A$, form
delta-function type kernels in the variable $y$ with two entries
$a\rightarrow (a,y),\,A\to (A,y')$,
    \begin{eqnarray}
    R^a_{\;A}=R^{a}_{\;\,A}(\partial_y)\,\delta(y-y').   \label{3.14}
    \end{eqnarray}
Various components of $R^a_{\;\,A}(\partial_y)$ are either the
ultralocal (multiplication) or differential operators acting on the
first argument of the delta function.

For sake of brevity, when using the condensed notations of the
canonical or spacetime nature we will not introduce special labels
to distinguish between them. As a rule, when the $y$-argument is
explicitly written we imply that the corresponding condensed indices
are canonical, i.e. they contain only discrete labels and brane
coordinates $x$, and their contraction does not involve implicit
$y$-integration. For example, the left-hand side of (\ref{3.3}) can
be written down in the form
    \begin{eqnarray}
    R^{\,\;\;a}_A\,\frac{\delta S^{\rm B}}
    {\delta G^a}=\stackrel{\rightarrow}{{\vphantom W}R}
    {\!\!\vphantom R}^{\;\;\;a}_A(\partial_y)\,
    \frac{\delta S}{\delta G^a(y)},
    \,\,\,A\rightarrow (A,y),                  \label{3.15}
    \end{eqnarray}
where the integration over $y$ (implicit in the contraction of the
{\it spacetime condensed} index $a$) removed the delta function
contained in $R^{\;\;\;a}_A$ and the result boiled down to the
action of the differential operator $R^{\;\;\;a}_A(\partial_y)$ on
$\delta S/\delta G^a(y)$. This operator obviously differs from that
of Eqs.(\ref{3.2}) and (\ref{3.14}) by the functional transpositon
-- integration by parts, because in contrast to (\ref{3.2}) it acts
on the test function with respect to the upper index $a$. This fact
is indicated by the order of operator indices reversed relative to
Eq. (\ref{3.14}).

Another distinction between these two types of condensed notations
concerns functional derivatives. We shall always reserve the
functional variational notation $\delta/\delta
G^a\equiv\delta/\delta G^a(y)$ for the variational derivative with
respect to the functions of $y$, while the variational derivative
with respect to the functions of brane coordinates will be denoted
by partial derivatives. For example, $\delta/\delta
G^a\equiv\delta/\delta G_{AB}(X)$ vs $\partial/\partial
g^i\equiv\delta/\delta g_{\alpha\beta}(x)$.

In these notations the action has the form
    \begin{eqnarray}
    S^{\rm B}[\,G\,]=
    \int dy\,L^{\rm B}\big(g,\dot{g},N\big) \label{3.16}
    \end{eqnarray}
where the ADM Lagrangian is displayed explicitly depending on $\dot
g^i=\dot G^i$, and all $x$-derivatives are implicit in condensed
canonical notations. With the definition of the momenta conjugated
to $g^i$
    \begin{eqnarray}
    p_{\,i}=p_i^0(g,\dot{g},N)
    \equiv\frac{\partial L^{\rm B}(g,\dot{g},N)}
    {\partial\dot{g}^{\,i}}                           \label{3.17}
    \end{eqnarray}
the bulk action can be rewritten in the canonical form
    \begin{eqnarray}
    &&S^{\rm B}[\,g,p,N\,]=\int dy\,
    \big(\, p_{\,i}\,\dot{g}^{\,i}-
    N^A \mbox{\boldmath$H$}_{\!A} (g,p)\big),        \label{999}\\
    &&S^{\rm B}[\,g,p,N\,]
    \big|_{\,p=p^0(g,\dot{g},N)}
    =S^{\rm B}[\,G\,],                              \label{1000}
    \end{eqnarray}
where $\mbox{\boldmath$H$}_{\!A}(g,p)=
(\mbox{\boldmath$H$}_{\!\alpha}(g,p),\mbox{\boldmath$H$}_{\!\perp}(g,p))$
is a set of momentum and Hamiltonian constraints. Similarly to the
ADM formalism in the physical time, these constraints as functions
on the phase space of $(g^i,p_i)$ satisfy the Poisson bracket
relations
    \begin{eqnarray}
    \{\,\mbox{\boldmath$H$}_{\!A},\mbox{\boldmath$H$}_{\!B}\}
    =U^C_{AB}\,\mbox{\boldmath$H$}_{\!C}             \label{3.18}
    \end{eqnarray}
with structure functions $U^C_{AB}=U^C_{AB}(g)$, signifying that
these constraints belong to the first class according to the Dirac
classification. As a consequence of (\ref{1000}) they also comprise
the $(\perp\! A)$-projections of the Einstein equations in the bulk,
    \begin{eqnarray}
    \mbox{\boldmath$H$}_{\!A} (g,p)\,
    \big|_{\,p=p^0(g,\dot{g},N)}
    =-\frac{\delta S^{\rm B}[\,g,N\,]}
    {\delta N^A}\,.                              \label{3.19}
    \end{eqnarray}

Due to the constraint algebra the canonical action (\ref{999}) is
invariant under the gauge transformations with local (arbitrary time
and space dependent) parameters $\Xi^A(X)$ satisfying
$\Xi^\perp(X)|=0$.\footnote{Tangential to the brane components
$\Xi^\mu|=\xi^\mu$ should not necessarily vanish because these
diffeomorphisms do not shift the boundary. In the canonical
formalism this property follows from the linearity of the momentum
constraints $H_\mu$ in $p$, due to which relevant surface terms
identically vanish for arbitrary $\Xi^\mu|$.} These transformations
are canonical and, therefore, ultralocal in $y$ for phase space
variables, but involve the $y$-derivative of $\Xi^A(X)$ for Lagrange
multipliers \cite{FV}
    \begin{eqnarray}
    &&\delta^\Xi g^{\,i}=
    \{\,g^{\,i},\mbox{\boldmath$H$}_{\!A}\}\;\Xi^A,\,\,\,
    \delta^\Xi p_{\,i}=\{\,p_{\,i},
    \mbox{\boldmath$H$}_{\!A}\}\;\Xi^A,    \label{3.20}\\
    &&\delta^\Xi N^A=\dot{\Xi}^A
    -U^A_{BC}\,N^B\, \Xi^C.             \label{3.21}
    \end{eqnarray}

In view of the relation (\ref{1000}) between the canonical and
Lagrangian formalisms various components of $R^a_{\;A}$ follow from
the transformations (\ref{3.20})-(\ref{3.21}) \cite{FV}
    \begin{eqnarray}
    &&R^i_A=\delta(y-y')\left.
    \frac{\partial\mbox{\boldmath$H$}_{\!A}}
    {\partial p_i}
    \,\right|_{\,p=p^0(g,\dot{g},N)},           \label{3.22}\\
    &&\nonumber\\
    &&R^B_{\;A}=\,\,\stackrel{\rightarrow}
    {{\vphantom W}R}{\!\vphantom R}^B_{\;A}
    (\partial_y)\,\delta(y-y')=\left(\delta^B_A\,\partial_y
    -U^B_{CA}\,
    N^C\right)\delta(y-y').             \label{3.23}
    \end{eqnarray}
The distinguished role of the Lagrange multiplyers manifests itself
in the fact that only the $a=B$ component of (\ref{3.14}) forms the
first order differential operator while the $a=i$ components are
ultralocal in $y$.

Sometimes a composition of differential operators results in an
ultralocal operator. Here is one important example that follows from
the transformation property of the momentum. On the one hand it is
given by the canonical transformation
    \begin{eqnarray}
    \delta^\Xi p_i=
    -\frac{\partial\mbox{\boldmath$H$}_{\!A}}
    {\partial g^i}\;\Xi^A.                         \label{3.24}
    \end{eqnarray}
On the other hand it can be obtained by the gauge transformation of
the Lagrangian expression for the momentum (\ref{3.17}). The metric
variation of the latter has the form of the differential operator
acting on $\delta G^a(y)=(\delta g^i(y),\delta N^A(y))$,
    \begin{eqnarray}
    \delta_G
    \frac{\partial L^{\rm B}(g,\dot{g},N)}
    {\partial\dot{g}^{\,i}}\equiv\,\,
    W^{\,\rm S}_{ia}(\partial)\,
    \delta G^a(y).                               \label{3.25}
    \end{eqnarray}
In essence $W^{\rm S}_{ia}(\partial)$ here is a part of the
Wronskian operator (\ref{2.21})) associated with the part of the
operator (\ref{2.13}), $F_{ab}=F^{AB,CD}(\nabla)\,\delta(X,X')$,
without the gauge-breaking term. Substituting the gauge variation of
the metric one therefore has
    \begin{eqnarray}
    \delta^\Xi p^0_i(g,\dot g,N)=\,\,
    \stackrel{\rightarrow}{{\vphantom W}W}
    {\!\!\vphantom W}^{\,\rm S}_{ia}\;
    \delta^\Xi G^a(y)=\,\,
    \stackrel{\rightarrow}{{\vphantom W}W}
    {\!\!\vphantom W}^{\,\rm S}_{ia}
    \stackrel{\rightarrow}{{\vphantom W}R}
    {\!\vphantom R}^a_{\,A}
    \,\Xi^A                                \label{3.26}
    \end{eqnarray}
and comparing (\ref{3.24}) and (\ref{3.26}) finds that the
composition of the two  first-order differential operators is
ultralocal in $y$ (and contains at most the derivatives with respect
to brane coordinates)\footnote{The equivalence of the canonical and
Lagrangian gauge transformations of momenta, $\big(\delta^\Xi
p\,\big)|_{\,p^0(g,\dot{g},N)}=
\delta^\Xi\big(p^0(g,\dot{g},N)\big)$, holds as it follows from
(\ref{1000}) only up to terms proportional to equations of motion
\cite{FV}. In what follows the bulk background always satisfies
on-shell condition which justifies the identification of canonical
and Lagrangian versions of gauge transformations.}
    \begin{eqnarray}
    \stackrel{\rightarrow}{W}
    {\!\!\vphantom W}^{\,\rm S}_{ia}(\partial)\!
    \stackrel{\rightarrow}{R}
    {\!\vphantom R}^a_{\,A}(\partial)=
    -\left.\frac{\partial \mbox{\boldmath$H$}_{\!A}}
    {\partial g^i}\,
    \right|_{\,p=p^0(g,\dot g,N)}                   \label{3.27}
    \end{eqnarray}

The final comment of this section concerns the diffeomorphisms
tangential to the brane. In the set of
$R^i_A=(R^i_\mu,\,\,R^i_\perp)$ their generators
$R^i_\mu=R^i_\mu(\,g\,)$ are independent of $\dot g^i$ and
correspond to the momentum constraints linear in $p$,
    \begin{eqnarray}
    \mbox{\boldmath$H$}_{\!\mu} (g,p)
    =R^i_\mu\,p_{\,i}.                       \label{3.28}
    \end{eqnarray}
By construction they leave the brane part of the action invariant
    \begin{eqnarray}
    R^i_\mu\,\frac{\partial S^{\rm b}}{\partial g^i}=0,  \label{3.29}
    \end{eqnarray}
which is the brane, $d$-dimensional, analogue of (\ref{3.3}).

\section{Gauge-fixing procedure}
The relativistic nature of the bulk gauge conditions (\ref{2.1})
which in condensed notations we denote by
    \begin{eqnarray}
    H^A=H^A\big(G^a(y),\dot G^a(y)\big)  \label{4.1}
    \end{eqnarray}
implies that they necessarily depend on the ``velocities" of the
Lagrange multipliers $\dot N^A=\partial_y N^A$, so that the
following matrix is nondegenerate
    \begin{eqnarray}
    a^A_B=-\frac{\partial H^A}{\partial\dot N^B},
    \,\,\,\,\det a^A_B\neq 0.                     \label{4.2}
    \end{eqnarray}
As a result the corresponding Faddeev-Popov operator in view of the
derivative nature of the transformations (\ref{3.21}) for $N^A$ is
of the second order in $\partial_y$. This follows from the
definition of this operator
    \begin{eqnarray}
    &&Q^A_B=\frac{\delta H^A}{\delta G^a}\, R^a_B=\,
    \stackrel{\rightarrow}{{\vphantom W}H}{\!\!\vphantom H}^A_{\;\;a}
    (\partial)
    \stackrel{\rightarrow}{{\vphantom W}R}
    {\!\!\vphantom R}^{\,a}_{\;\;B}
    (\partial)\,\delta(y-y'),               \label{4.3}\\
    &&\frac{\delta H^A}{\delta G^a}
    \equiv H^A_{\;\;a}=\,
    \stackrel{\rightarrow}{{\vphantom W}H}{\!\!
    \vphantom H}^A_{\;\,a}
    (\partial)\,\delta(y-y'),           \label{4.4}
    \end{eqnarray}
where $H^A_{\;\;a}$ is the matrix of linearized gauge conditions
(\ref{2.16})). This functional matrix in relativistic gauges is the
first-order differential operator,
$H^A_{\;\;C}(\partial_y)=-a^A_C\partial_y+...$, whence in view of
(\ref{3.23})
    $Q^A_{\;\,B}(\partial_y)=-a^A_B\,
    \partial_y^2+...\;.$
Such a second-order operator admits zero modes parameterized by
their boundary conditions, corresponding to the residual
transformations discussed in Introduction and in Sect.2.

To fix these residual transformations we impose gauge conditions on
the brane metric (\ref{2.2}), $\chi^\mu(g)=0$. They force these
boundary conditions to vanish, provided the brane Faddeev-Popov
operator, which in condensed canonical notations reads as
    \begin{eqnarray}
    \mbox{\boldmath$J$}^\mu_\nu=
    \frac{\partial\chi^\mu}{\partial g^i}\, R^i_\mu,        \label{4.8}
    \end{eqnarray}
is nondegenerate. As discussed in Introduction, the brane gauge
conditions can be chosen background-covariant from the
$d$-dimensional viewpoint, as is the case of the brane DeWitt gauge
(\ref{4.9}). However, they are imposed only on
$g^i=g_{\alpha\beta}(x)$ and do not involve $\dot g^i$, so that they
can be considered unitary from the viewpoint of the $y$-time
canonical formalism.

Thus, the overall gauge-fixing factor in the path integral
(\ref{1.2}) takes the form
    \begin{eqnarray}
    (\,\mbox{gauge-fixing}\,)
    =\delta\left[\,H^A(\,G\,)\,\right]\,
    \delta(\chi^\mu(g))\,M_{\,H,\;\chi}[\,G\,],    \label{4.11}
    \end{eqnarray}
where we use different condensed notations for the functional
delta-functions in the bulk (\ref{2.6}) and on the brane
(\ref{2.7}). The measure factor $M_{\,H,\;\chi}[\,G\,]$ is
determined according to the standard Faddeev-Popov procedure by the
following functional integral over the full diffeomorphism group
    \begin{eqnarray}
    \left(\,M_{\,\chi,\,H\,}[\,G\,]\,\right)^{-1}=
    \int\limits_{\footnotesize
    \begin{array}{c}
    \Xi^\perp|\,=0\\
     \end{array}}
    \!\!\!\!D\,\Xi\;\delta
    \left[\,H^A(\,G_\Xi\,)\,\right]\,
    \delta(\chi^\mu(\,g_\xi\,)).                \label{4.12}
    \end{eqnarray}
To calculate it we write the infinitesimally transformed bulk and
brane metrics as $G^a_\Xi=G^a+R^a_A\,\Xi^A+O[\;\Xi^2\,]$ and
$g^i_\xi=g^i+R^i_\mu\,\xi^\mu+O[\;\xi^2\,]$, and then decompose in
(\ref{4.12}) the integration over $\Xi^A(X)$ into the integration
over the bulk field with fixed boundary values
$\Xi^A(X)\,|=(0,\xi^\mu(x))$ and the subsequent integration over
$\xi^\mu(x)$
    \begin{eqnarray}
    &&\int\limits_{\footnotesize
    \begin{array}{c}
    \Xi^\perp|\,=0\\
     \end{array}}
    \!\!\!\!D\,\Xi\;\delta\left[\,H^A(\,G_\Xi\,)\,\right]\,
    \delta(\chi^\mu(\,g_\xi\,))=
    \int d\xi \,\delta\!\left(\,
    \mbox{\boldmath$J$}^\mu_\nu\xi^\nu\,\right)
    \!\!\!
    \int\limits_{\footnotesize
    \begin{array}{c}
    \Xi^\perp|\,=0\\
    \Xi^\mu|\,=\xi^\mu
     \end{array}}
    \!\!\!\!D\,\Xi\;\delta\!\left[\;Q^A_B\,\Xi^B\,\right]\,
    \nonumber\\
    &&\nonumber\\
    &&\qquad\qquad=\left(\,{\rm det}\,
    \mbox{\boldmath$J$}^\mu_\nu\,\right)^{-1}
    \!\!\!\!\!\int\limits_{\,\,\,\,\,\,\,\,\,\,\footnotesize
    \begin{array}{c}
    \Xi^A|\,=0\\
     \end{array}}
    \!\!\!\!\!\!D\,\Xi\;\delta\!\left[\;Q^A_B\,\Xi^B\,\right]
    =\left(\,{\rm det}\,
    \mbox{\boldmath$J$}^\mu_\nu\,\right)^{-1}
    \left(\,{\rm Det}_{\rm D}\,Q^A_B\,\right)^{-1}.  \label{4.13}
    \end{eqnarray}
The result is structurally very simple -- it factorizes into the
product of functional determinants of brane and bulk Faddeev-Popov
operators, and the latter, ${\rm Det}_{\rm D}\,Q^A_B$, is calculated
subject to Dirichlet conditions for {\em all} components of
$\Xi^A|$. This directly follows from the gauge-fixing procedure on
the brane.

The Dirichlet type functional determinant is determined by the
variational formula
    \begin{eqnarray}
    \delta\,{\rm ln}\,{\rm Det}_{\rm D}\,Q^A_B=
    Q^{-1\,B}_{\,{\rm D}\;A}\!\stackrel{\leftarrow}
    {\delta Q}{\!\!}^{\;\;A}_{B}\equiv\int dy\;
    Q^{-1\;\;B}_{\,{\rm D}\,A}(y',y)
    \stackrel{\leftarrow}
    {\delta Q}{\!\!}^{\;\;A}_{B}(\partial_y)
    \Big|_{\,y'=y},                                    \label{4.13a}
    \end{eqnarray}
where $\stackrel{\leftarrow}{\delta Q}{\!\!}^{\;\;A}_{B}$ denotes
the variation of the operator under generic change of its
coefficients and $Q^{-1 B}_{\,{\rm D}\;A}$ is the Dirichlet Green's
function of the ghost operator, defined by
    \begin{eqnarray}
    &&Q^{C}_{\;\;B}(\partial_y)\,
    Q^{-1 B}_{\;{\rm D}\;\;\;A}(y,y')
    =\delta^C_A\,\delta(y-y'),    \nonumber\\
    &&Q^{-1 B}_{\;{\rm D}\;\;\;A}(y,y')\big|_{\,y}=0.     \label{4.13b}
    \end{eqnarray}
Thus, this derivation yields one of the main results of this paper
(\ref{2.5}) and confirms boundary conditions for the ghost
propagator in the bulk.

As a result of transition (\ref{2.8})-(\ref{2.10a}) the quantum
effective action takes the form
    \begin{eqnarray}
    &&e^{\textstyle i\SGamma}=\int DG\,\mu[\,G\,]\,
    \exp\Big(\,i S_{\rm gf}[\,G\,]\Big)\,
    \det \mbox{\boldmath$J$}^\mu_\nu\;
    {\rm Det}_{\rm D} Q^A_B,           \label{4.14}\\
    &&S_{\rm gf}[\,G\,]=S^{\rm B}[\,G\,]
    +S^{\rm B}_{\rm gb}[\,G\,]+S^{\rm b}[\,g\,]
    +S^{\rm b}_{\rm gb}[\,g\,],                    \label{4.15}
    \end{eqnarray}
with the full gauge-fixed action $S_{\rm gf}[\,G\,]$ including the
bulk and brane gauge-breaking terms which read in canonical
condensed notations as
    \begin{eqnarray}
    &&S^{\rm B}_{\rm gb}[\,G\,]=-\frac 12\,
    \int dy\, H^A(G,\dot G)\,
    c_{AB}\,H^B(G,\dot G),\\
    &&S^{\rm b}_{\rm gb}(g)=-\frac 12\,
    \chi^\mu(g)\,c_{\mu\nu}\chi^\nu(g).                 \label{4.16}
    \end{eqnarray}

The last comment of this section concerns the local measure denoted
in (\ref{4.14}) by $\mu[\,G\,]$. Its contribution is not very
important for practical purposes, because its function solely
consists in the cancellation of strongest power divergences of the
path integral. Nevertheless, for completeness we briefly discuss it
here.

The local measure is determined by another type of bulk and brane
foliation -- the one with the slices of constant physical time
$t=X^0$. This foliation determines the chronological ordering in the
unitary evolution in physical (not fictitious) time. It gives rise
to the local measure as a contribution of the gaussian path integral
over the physical momenta conjugated to temporal velocities
$\partial_0 G$ (rather than to $\partial_y G$). As shown in
\ref{Measure}, when the brane action has the kinetic
term\footnote{Which is not always the case -- in the Randall-Sundrum
model, for example, the brane part of the action is kinetically
inert, because it contains only the brane tension.}, the full local
measure factorizes into the product of the bulk and brane measures
    \begin{eqnarray}
    \mu[\,G\,]=\mu_{\rm B}[\,G\,]\,
    \mu_{\rm b}(\,g\,).                            \label{4.17}
    \end{eqnarray}

The bulk measure looks as follows. In the foliation of the bulk by
spacelike slices of constant $X^0=t$, $X=(t,{\bf x},y)$ (where $\bf
x$ denotes spatial coordinates among brane coordinates $x$), the
bulk metric can be decomposed into the corresponding spacelike
metric $Q(t,{\bf x},y)$ and bulk lapse and shift functions ${\cal
N}(t,{\bf x},y)\sim G_{At}(X)$, $G_{AB}(X)\sim(Q(t,{\bf x},y),{\cal
N}(t,{\bf x},y))$. The bulk measure then reads as
    \begin{eqnarray}
    \mu_{\rm B}[\,G\,]=\prod_{X,X\not\in {\rm\bf b}}\,
    \left[\,{\rm det}\frac{\partial^2 L_B(Q,\partial_tQ,{\cal N})}
    {\partial(\partial_t Q)\,
    \partial(\partial_t Q)}\;
    {\rm det}\,c_{AB}\right]^{1/2}.                 \label{4.18}
    \end{eqnarray}
It also absorbs the determinant of the gauge-fixing matrix $c_{AB}$
which is generated by the transition (\ref{2.8})-(\ref{2.10a}) to
nondegenerate gauges and which we also consider ultralocal in $X$.

Similarly, in the construction of the brane measure one has a
foliation of the brane by spacelike $(d-1)$-dimensional surfaces
$x=(t,{\bf x})$, which leads to $\big((d-1)+1\big)$-decomposition
$g(x)=(q(t,{\bf x}),n(t,{\bf x}))$ into the space metric $q(t,{\bf
x})$ and brane lapse and shift functions $n(t,{\bf x})$. The brane
Lagrangian then depends on velocities of only the $q$-metric
coefficients, $L_{\rm b}(g,\partial g)=L_{\rm b}(q,\partial_t q,n)$,
and the brane measure takes the form
    \begin{eqnarray}
    &&\mu_{\rm b}(\,g\,)=\prod_{x}\,\left[\,{\rm det}
    \frac{\partial^2 L_b}
    {\partial(\partial_t q)\,
    \partial(\partial_t q)}\;
    {\rm det}\,c_{\mu\nu}\right]^{1/2}.                  \label{4.19}
    \end{eqnarray}

\section{Bulk wavefunction representation of the brane effective action}
As a next step we decompose the full integration in (\ref{4.14})
into the integration over the bulk metric $G_{AB}(X)$ subject to
fixed induced metric on the brane
$g_{\alpha\beta}(x)=G_{\alpha\beta}(X)\,|$ and the subsequent
integration over $d$-dimensional $g_{\alpha\beta}(x)$,
    \begin{equation}
    \int DG_{AB}(X)=\int
    dg_{\alpha\beta}(x)\!\!\!\!\!\!\!
    \int\limits_{\,\,\,\,\,\,\,
    G_{\alpha\beta}\,|\,=\,g_{\alpha\beta}(x)}
     \!\!\!\!\!\!\!\!\!\!\!DG_{AB}(X)
     \equiv
     \int dg \int\limits_{\,G^i|\,=\,g^i}\!
    \!\!\!DG.                           \label{5.1}
    \end{equation}
The result looks as the Feynman-DeWitt-Faddeev-Popov functional
integral \cite{gospel}
    \begin{equation}
    e^{\textstyle i\SGamma}=\int
    dg\;\mu_{\,\rm b}[\,g\,]\,
    \exp i\!\left( S^{\rm b}(\,g\,)+
    S^{\rm b}_{\rm gb}(\,g\,)\right)\,
    \det \mbox{\boldmath$J$}^\mu_\nu\;
    \mbox{\boldmath$\varPsi$}_{\!\rm B}(\,g\,)     \label{5.2}
    \end{equation}
for the purely $d$-dimensional system with the brane action $S^{\rm
b}[\,g_{\alpha\beta}\,]$ but with the insertion of the functional
$\mbox{\boldmath$\varPsi$}_{\!\rm B}(\,g\,)$ which we will call a
wavefunction of the bulk spacetime
    \begin{equation}
    \mbox{\boldmath$\varPsi$}_{\!\rm B}(\,g\,)
    =\int\limits_{\,G^i|\,=\,g^i}\!
    \!\!\!DG\,\mu_{\rm B}[\,G\,]\,
    \exp i\!\left( S^{\rm B}[\,G\,]
    + S^{\rm B}_{\rm gb}[\,G\,]\,\right)\;
    {\rm Det}_{\rm D}\, Q^A_B.                      \label{5.3}
    \end{equation}

This function is well known from quantum cosmology as a
path-integral representation of the solution of the Wheeler-DeWitt
equations -- quantum Dirac constraints on a quantum state in the
canonical quantization of gravity. At various levels of rigor and in
various contexts it was built in a path-integral form in
\cite{Leutw,HH,barvin,HalHar,Dirac}. In particular, in the $d=3$
context it was obtained in \cite{barvin} where the initial and final
spacelike Cauchy surfaces played the role of branes with specified
3-metrics. The metrics served as functional arguments of the
two-point kernel of the unitary evolution interpolating between
these two 3-surfaces.

In (\ref{5.3}) the role of the boundary is played by timelike
brane(s), and the canonical formalism in real time is generalized to
the case of the ``evolution" in the direction transversal to the
brane. Certainly, no sensible unitarity or causality can be ascribed
to this evolution. Apparently this construction can be generalized
to a purely Euclidean case when the full boundary can have various
topological and connectedness properties.

The bulk wavefunction (\ref{5.3}) has the same properties as the
cosmological wavefunction of \cite{barvin,Dirac}. First of all, it
is independent of the choice of the bulk gauge $H^A$,
    \begin{equation}
    \delta_H \mbox{\boldmath$\varPsi$}_{\!\rm B}(\,g\,)=0, \label{5.4}
    \end{equation}
though $H^A$ explicitly enters its construction. This a typical
property of the Faddeev-Popov functional integral on shell (or in
the absence of sources located in the bulk --- the only source of
$\mbox{\boldmath$\varPsi$}_{\!\rm B}(\,g\,)$ is the brane metric
$g=g_{\alpha\beta}(x)$ ). A formal proof is based on the change of
the integration variable in (\ref{5.3})
    \begin{equation}
    G^a\to G^a_{\Xi}=G^a+R^a_A\,\Xi^A,\,\,\,\,
    \Xi^A=Q^{-1\,A}_{\;B}\,\delta H^B,                   \label{5.5}
    \end{equation}
simulating the change of gauge conditions $H^A\to H^A+\delta H^A$.
Due to Ward identities for ghost and gauge Green's functions and
gauge invariance of the bulk action, this transformation is
identical and leads to (\ref{5.4}). Important point of the transform
(\ref{5.5}) is that the Green's function of the Faddeev-Popov
operator $Q^{-1\,A}_{\;B}$ has Dirichlet boundary conditions, and
therefore $\Xi^A|=0$, so that this transform does not shift $g^i$ on
the brane (as it should). It only shifts lapse and shift functions
$N^C|$ because of the $y$-derivative in $R^C_A(\partial_y)$ acting
on $\Xi^A(y)$, but this is not dangerous because the Lagrange
multiplyers $N^C|$ are integrated over at the boundary, rather than
being fixed like $G^i|=g^i$.\footnote{This is a point of departure
from \cite{scale} where the effective action is a functional of
$N^C|$ and, therefore, depends on the choice of the gauge in the
bulk.}

Important consequence of this integration over $N^C|$ is that
$\mbox{\boldmath$\varPsi$}_{\!\rm B}(\,g\,)$ satisfies the analogue
of the Wheeler-DeWitt equations
    \begin{equation}
    \hat{\!\mbox{\boldmath$H$}}_{\!A}
    (g,\partial/i\partial g)\,
    \mbox{\boldmath$\varPsi$}_{\!\rm B}(\,g\,)=0,       \label{5.6a}
    \end{equation}
where $\hat{\!\mbox{\boldmath$H$}}_{\!A}(g,\partial/i\partial g)$
are the operators of quantum Dirac (momentum and Hamiltonian)
constraints. They follow from their classical counterparts by some
nontrivial operator realization which is formally known only in the
linear in $\hbar$ approximation
\cite{operator}.\footnote{\label{footnote}It is not obvious, though,
that the realization of \cite{operator} for the temporal constraints
would work for brane constraints, because the latter govern an
artificial evolution in the $y$-time which is not related to
unitarity and causality. In particular, the choice of the local
measure in (\ref{5.3}) is not related to $y$-foliation of the bulk
and does not yield the scalar density nature of
$\mbox{\boldmath$\varPsi$}_{\!\rm B}(\,g\,)$ under the
reparameterizations of $g$. This choice is tightly related to the
operator realization of quantum Dirac constraints. For the temporal
constraints of  \cite{operator} the choice of measure guarantees
that the cosmological wavefunction is a scalar density of
$1/2$-weight, and the operators
$\hat{\!\mbox{\boldmath$H$}}_{\!A}(g,\partial/i\partial g)$ act
covariantly on this density. No such properties hold for
$\mbox{\boldmath$\varPsi$}_{\!\rm B}(\,g\,)$ -- a scalar in $g$
rather than a density, see \ref{Measure}.}

The momentum components of (\ref{5.6}), $A=\mu$, describe the gauge
invariance of $\mbox{\boldmath$\varPsi$}_{\!\rm B}(\,g\,)$ under
$d$-dimensional diffeomorphisms. In condensed notations this reads
as
    \begin{eqnarray}
    R^i_\mu\,\frac{\partial
    \mbox{\boldmath$\varPsi$}_{\!\rm B}}
    {\partial g^i}=0.                      \label{5.6}
    \end{eqnarray}
This property can also be independently proven by the transformation
of integration variable similar to (\ref{5.5}) but with
$\Xi^\mu|=\xi^\mu\neq 0$ -- the diffeomorphism tangential to the
brane \cite{HalHar}. This property in its turn guarantees the
$\chi$-gauge independence of the {\em on-shell} brane action
$\varGamma$,
    \begin{equation}
    \delta_\chi \varGamma=0.    \label{5.7}
    \end{equation}
Indeed, the $d$-dimensional analogue of the change of integration
variables (\ref{5.5}) in the path integral (\ref{5.2}) gives this
property in virtue of gauge invariance of $S^{\rm b}(\,g\,)$ and
$\mbox{\boldmath$\varPsi$}_{\!\rm B}(\,g\,)$ and the Ward identities
for the brane Faddeev-Popov Green's function.

\section{Semiclassical expansion}
\subsection{Wavefunction of the bulk}
Semiclassical expansion for $\mbox{\boldmath$\varPsi$}_{\!\rm
B}(\,g\,)$ is based on the stationary point of the gauge-fixed
action in the bulk. As a result of integration by parts, the
variation of the action includes the bulk and boundary terms
    \begin{eqnarray}
    &&\delta\left( S^{\rm B}[\,G\,]+
    S^{\rm B}_{\rm gb}[\,G\,]\,\right)
    =\int\limits_B dX\,\Big(\frac{\delta S^{\rm B}}{\delta G^a(X)}
    -\frac{\stackrel{\rightarrow}{\delta H}
    {\!\!\vphantom H}^D}{\delta G^a(X)}\,c_{DB}H^B\Big)
    \delta G^a(X)\nonumber\\
    &&\qquad\qquad\qquad\qquad\qquad\qquad
    +\left.\Big(\,\frac{\partial L_{\rm B}}{\partial\dot G^a}
    -\frac{\partial H^D}
    {\partial\dot G^a}\,c_{DB}H^B\Big)
    \delta G^a\right|                        \label{6.1}
    \end{eqnarray}
which separately should be equated to zero. Now, take into account
that for fixed $G^i|=g^i$ the boundary variation $\delta G^i|=0$ and
note that $L_B$ is independent of $\dot N^A$ and the matrices
$\partial H^D/\partial\dot N^A=-a^D_A$ (cf. Eq.(\ref{4.2})) and
$c_{DB}$ are invertible. Therefore, the stationarity conditions
reduce to
    \begin{eqnarray}
    &&\frac{\delta S^{\rm B}}{\delta G^a(X)}
    -\frac{\delta H^D}{\delta G^a(X)}
    \,c_{DB}H^B=0,                       \label{6.2}\\
    &&H^B|=0.                             \label{6.3}
    \end{eqnarray}
Functionally contracting (\ref{6.2}) with $R^a_A$ and using gauge
invariance of the bulk action (\ref{3.3}) we get for the gauge
conditions the homogeneous equation with the second-order
Faddeev-Popov operator
    \begin{eqnarray}
    Q^D_A(\partial)\,c_{DB}H^B(X)=0.           \label{6.4}
    \end{eqnarray}
Since in view of (\ref{6.3}) they have Dirichlet boundary
conditions, the solution is identically zero everywhere in the bulk.
Thus, eventually the stationary point of the bulk action
    \begin{eqnarray}
    G_0^a(\,g\,)\equiv
    G^0_{AB}(X)[\,g_{\alpha\beta}(x)\,]  \label{6.5}
    \end{eqnarray}
is a solution of classical equations of motion in the bulk gauge
with fixed brane metric at the boundary, as advocated in
Eqs.(\ref{2.12})-(\ref{2.12b}) of Sect.2.

The quadratic part of the action on this background -- the variation
of (\ref{6.1}) -- reads as
    \begin{equation}
    \frac12\,\delta^2\big(S^{\rm B}[\,G\,]
    + S^{\rm B}_{\rm gb}[\,g\,]\big)=
    \frac12\,\delta G^a \left(
    \stackrel{\rightarrow}{{\vphantom F}F}_{ab}\!\delta G^b\right)
    +\frac12\,\delta G^a \!
    \stackrel{\rightarrow}
    {{\vphantom W}W}_{ab}\!(\partial)\,
    \delta G^b\Big|\,,               \label{6.6a}
    \end{equation}
where $\stackrel{\rightarrow}{F}_{ab}$ is the condensed notation for
the operator (\ref{2.13})
    \begin{eqnarray}
    &&F_{ab}=S_{ab}-
    H^A_a\,c_{AB}H^B_{\,\,b},             \label{6.7a}\\
    &&S_{ab}\equiv\,
    \stackrel{\rightarrow}{S}_{ab}\!(\partial_y)\,\delta(y-y')=
    \frac{\,\delta^2 S^{\rm B}[\,G\,]}
    {\delta G^a(y)\,\delta G^b(y')}                    \label{6.8a}
    \end{eqnarray}
and $\stackrel{\rightarrow}{W}_{ab}\!(\partial)$ is the
corresponding Wronskian operator (\ref{2.21}) defined by the
variational relation
    \begin{eqnarray}
    &&\stackrel{\rightarrow}{{\vphantom W}W}_{ab}\!(\partial)\,
    \delta G^b(y)=
    \delta\!\left(\,\frac{\partial L^{\rm B}}{\partial\dot G^a}
    -\frac{\partial H^A}{\partial\dot G^a}
    \,c_{AB}H^B\right)_{H^A=0}\nonumber\\
    &&\qquad\qquad\qquad\qquad\qquad\qquad
    =\,\Big(\!\stackrel{\rightarrow}{{\vphantom W}W}
    {\!\!\vphantom W}^{\,\rm S}_{ab}(\partial)
    -\frac{\partial H^A}{\partial\dot G^a}\,c_{AB}\!
    \stackrel{\rightarrow}{{\vphantom W}H}
    {\!\!\vphantom H}^B_{\,b}(\partial)\Big)\,
    \delta G^b(y).                            \label{6.9a}
    \end{eqnarray}
Together with the Wronskian operator of the invariant action
$W^{\,\rm S}_{ab}(\partial)$, see Eq. (\ref{3.25}), it includes the
contribution of the gauge-breaking term.
$\stackrel{\rightarrow}W_{ab}\!(\partial)$ participates in the
Wronskian relation (\ref{2.21}) and also satisfies an additional
relation that can be obtained by a single integration by parts of
the derivatives in $\stackrel{\rightarrow}{F}_{ab}\!(\partial)$
    \begin{equation}
      \phi^a_{1}\!
      \stackrel{\rightarrow}{F}_{ab}\!\phi^b_{2}=
      \phi^a_{1}\!\stackrel{\leftrightarrow}{F}_{ab}\!
      \phi^b_{2}-\phi^a_{1}\!
      \stackrel{\rightarrow}{W}_{ab}\!(\partial)\,
      \phi^b_{2}\,\Big|                             \label{6.10a}
      \end{equation}
Here $\stackrel{\leftrightarrow}{F}_{ab}$ implies that the
derivatives in $\stackrel{\rightarrow}{F}_{ab}(\partial)$ are
integrated by parts one time to form the bilinear combinations of
the first-order derivatives of $\phi_{1,2}$ (for $\phi_{2}=\phi_{1}$
this is just the Lagrangian quadratic in $\phi$ and $\partial\phi$).
With this relation the quadratic form of the action (\ref{6.6a})
takes the form
    \begin{equation}
    \frac12\,\delta^2\big(S^{\rm B}[\,G\,]
    + S^{\rm B}_{\rm gb}[\,g\,]\big)=
    \frac12\,\delta G^a \!
    \stackrel{\leftrightarrow}{{\vphantom F}F}_{ab}\!
    \delta G^b.                                        \label{6.6}
    \end{equation}

The one-loop contribution of this form to the path integral
(\ref{5.3}) is the gaussian functional integral over the
perturbations $\delta G^a=\delta G_{AB}(X)$ subject to Dirichlet
boundary conditions for the brane metric components $\delta G^i|=0$,
while $\delta N^A|$ are integrated over in the infinite limits. As
shown in \ref{DNGauss} this integral yields the functional
determinant of $F_{ab}$ subject to mixed Dirichlet-Neumann boundary
conditions, and the one-loop wavefunction of the bulk has the final
form
    \begin{eqnarray}
    \mbox{\boldmath$\varPsi$}_{\rm B}(\,g\,)= \left.
    \frac{\mu_{\rm B}\;{\rm Det}_{\rm D}\,Q^A_B}
    {\big(\,{\rm Det}_{\rm DN} F_{ab}\,\big)^{1/2}}\,
    \exp \Big( i S^{\rm B}[\,G\,]\Big)\,
    \right|_{\,G\,=\,G_0(g)}.                         \label{6.7}
    \end{eqnarray}
Here the Dirichlet-Neumann functional determinant ${\rm Det}_{\rm
DN} F_{ab}$ is determined by the variational relation
    \begin{eqnarray}
    \delta\,{\rm ln}\,{\rm Det}_{\rm DN}F_{ab}\equiv
    \delta\,{\rm Tr}_{\rm DN}{\rm ln}\,F_{ab}=\,\,
    \stackrel{\leftrightarrow}
    {\delta F}{\!\!}_{ab}\,G^{ba}_{\,\,\rm DN}.        \label{6.8}
    \end{eqnarray}
The notation $\stackrel{\leftrightarrow}{\delta F}{\!\!}_{ab}$ means
arbitrary variations of the coefficients of the operator, and,
similarly to Eq.(\ref{6.10a}), the double arrow implies symmetric
action of two first-order derivatives of $\delta F_{ab}=\delta
F^{AB,CD}(\nabla)$ on both arguments of the Green's function $G_{\rm
DN}^{ba}=G^{\rm DN}_{CD,AB}(X,X')$ (before taking the coincidence
limit $X'=X$ implicit in the functional trace operation).

The most peculiar element of the variational definition (\ref{6.8})
is $G^{ba}_{\;\rm DN}$ -- the Green's function of the operator
$F_{ab}(\partial)$ subject to the mixed set of Dirichlet and
generalized Neumann boundary conditions (\ref{2.15})-(\ref{2.15b}).
In canonical condensed notations this boundary value problem reads
as
    \begin{eqnarray}
    &&F_{ca}(\partial) \,G^{ab}_{\,\,\rm DN}(y,y')=
    \delta^b_c\,\delta(y-y'),                          \label{6.9}\\
    &&G^{ib}_{\,\,\rm DN}(y,y')\big|_{\;y}=0,          \label{6.10}\\
    &&\stackrel{\rightarrow}{H}
    {\!\!\vphantom H}^A_{\,a}(\partial)\,
    G^{ab}_{\,\,\rm DN}(y,y')\big|_{\;y}=0.             \label{6.11}
    \end{eqnarray}
Thus, contrary to the purely Dirichlet ghost propagator the metric
propagator $G^{ab}_{\rm DN}$ has Dirichlet conditions only for
$i=(\alpha\beta,X)$ components. The rest of boundary conditions
(\ref{6.11}) belong to the generalized Neumann type. This is a
consequence of the fact that lapse and shift functions in the path
integral for $\mbox{\boldmath$\varPsi$}_{\rm B}(\,g\,)$ are
integrated out on the brane.

\subsection{Brane effective action}
Now substitute one-loop bulk wavefunction (\ref{6.7}) into the path
integral (\ref{5.2}) and calculate it by the stationary-phase
method. The stationary point of the overall tree-level phase
satisfies the following equation
    \begin{equation}
    \frac{\partial}{\partial g^i}\,
    \Big(S^{\rm B}[\,G_0(g)\,]+
    S^{\rm b}(\,g\,)+ S^{\rm b}_{\rm gb}(\,g\,)\Big)=
    \left.\frac{\partial L_{\rm B}}{\partial\dot G^i}\,\right|+
    \frac{\partial S^{\rm b}}{\partial g^i}
    -\frac{\partial\chi^\mu}{\partial g^i}
    \,c_{\mu\nu}\chi^\nu=0,                       \label{6.12}
    \end{equation}
because $S^{\rm B}[\,G_0(g)\,]$ plays the role of the
Hamilton-Jacobi function in the $y$-time canonical formalism, and
its gradient in $g^i$ yields the canonical momentum on the brane
$p_i^0(G,\dot G)=\partial L_{\rm B}/\partial\dot G^i$,
    \begin{eqnarray}
    \frac{\partial}{\partial g^i}\,S^{\rm B}[\,G_0(g)\,]=
    \left.\frac{\partial L_{\rm B}}{\partial\dot G^i}
    \,\right|.                                     \label{6.13}
    \end{eqnarray}

In essence (\ref{6.12}) represents the generalized Israel matching
condition relating the extrinsic curvature of the brane,
$K^{\alpha\beta}(x)\sim\partial L_{\rm B}/\partial\dot
G_{\alpha\beta}(x)$, to the brane stress tensor $\partial S_{\rm
b}/\partial g^i=\delta S_{\rm b}/\delta g_{\alpha\beta}(x)$ in the
presence of the brane gauge-breaking term. To handle the latter,
contract (\ref{6.12}) with $R^i_\mu$ and take into account that in
view of (\ref{3.19}) and (\ref{3.28}) the first term in the
contraction vanishes on bulk shell,
    \begin{eqnarray}
    R^i_\mu\,\frac{\partial L_B}{\partial\dot G^i}
    =-\left.\frac{\delta S^{\rm B}}
    {\delta N^\mu}\,\right|_{\;G=G_0(g)}\equiv 0,    \label{6.14}
    \end{eqnarray}
whereas the second term vanishes in view of $d$-dimensional
covariance of the brane action (\ref{3.29}), so that finally
    \begin{equation}
    R^i_\mu\frac{\partial\chi^\lambda}{\partial g^i}
    \,c_{\lambda\sigma}\,\chi^\sigma\equiv
    \mbox{\boldmath$J$}^\lambda_\sigma\,
    c_{\lambda\sigma} \chi^\sigma=0,                 \label{6.15}
    \end{equation}
whence $\chi^\mu=0$ due to invertibility of
$\mbox{\boldmath$J$}^\lambda_\sigma$ and $c_{\lambda\sigma}$.
Therefore, we get the following system of equations for the
stationary point $g_0$ in the $d$-dimensional brane gauge
    \begin{eqnarray}
    &&\left(\left.\frac{\partial L_B}
    {\partial\dot G^i}\,\right|+
    \frac{\partial S^{\rm b}}
    {\partial g^i}\right)_{g\,=\,g_0}=0,             \label{6.16a}\\
    &&\chi^\mu(\,g_0)=0.                            \label{6.16}
    \end{eqnarray}

To find the quadratic form of the tree-level action on the
background of $g_0$ we would need the derivative of $G^a_0(g)$ with
respect to $g^i$. This quantity satisfies the linearized version of
the boundary-value problem (\ref{2.12})-(\ref{2.12b}). In condensed
notations this problem for the bulk perturbation $\phi^a(\varphi)$
induced by the perturbation $\varphi^i$ on the boundary reads as
    \begin{eqnarray}
    &&\stackrel{\rightarrow}{S}_{ab}\!
    (\partial)\,\phi^b(y)=0,             \label{6.17}\\
    &&\stackrel{\rightarrow}{H}
    {\!\!\vphantom H}^A_{\,b}(\partial)
    \,\phi^b(y)=0,\,\,\,\,\phi^i\big|=\varphi^i.         \label{6.19}
    \end{eqnarray}
To solve it notice that the $Ab$ component of the Wronskian operator
(\ref{6.9}) is entirely determined by the gauge-breaking term of the
full action
    \begin{eqnarray}
    &&\stackrel{\rightarrow}{W}_{Ab}\!(\partial)
    =-\frac{\partial H^B}{\partial\dot N^A}
    \,c_{BD}\!
    \stackrel{\rightarrow}{H}
    {\!\!\vphantom H}^D_{\,b}(\partial)
    =a_{AD}\!\stackrel{\rightarrow}{H}
    {\!\!\vphantom H}^D_{\,b}(\partial),   \label{6.20}\\
    &&a_{AD}\equiv a^B_A\,c_{BD}.
    \end{eqnarray}
Therefore the problem (\ref{6.17})-(\ref{6.19}) can be identically
rewritten in the form
    \begin{eqnarray}
    &&\stackrel{\rightarrow}{F}_{ab}\!
    (\partial)\,\phi^b(y)=0,                   \label{6.21}\\
    &&\stackrel{\rightarrow}{W}_{Ab}\!
    (\partial)\,\phi^b\big|=0,
    \,\,\,\,\,\phi^i\big|=\varphi^i,        \label{6.23}
    \end{eqnarray}
where the nondegenerate operator $F_{ab}$ replaces the degenerate
operator $S_{ab}$ defined by (\ref{6.8a}). This is a problem with
the (inhomogeneous) Dirichlet-Neumann boundary conditions. By using
the Wronskian relation (\ref{2.22}) its solution $\phi^a_{\rm DN}$
can be presented in terms of the Dirichlet-Neumann Green's function
of the above type
    \begin{eqnarray}
    \phi^a_{\rm DN}(y)=-G^{ab}_{\,\,\rm DN}(y,y')\!
    \stackrel{\leftarrow}{{\vphantom F}W}_{bi}\!
    (\partial')\big|_{\,y'}\varphi^i\equiv
    -G^{ab}_{\,\,\rm DN}\!
    \stackrel{\leftarrow}
    {{\vphantom F}W}_{bi}\!\!\big|\;\varphi^i, \label{6.24}
    \end{eqnarray}
whence
    \begin{eqnarray}
    \frac{\partial G_0^a(g)}{\partial g^i}=
    -G^{ab}_{\rm DN}\!
    \stackrel{\leftarrow}{W}_{\!bi}\!\big|\;.    \label{6.25}
    \end{eqnarray}

Using (\ref{6.25}) together with (\ref{6.13}) and (\ref{3.25}) one
immediately obtains
    \begin{eqnarray}
    \frac{\partial^2 S^{\rm B}[\,G_0(g)\,]}
    {\partial g^i\,\partial g^k}\,
    =-\!
    \stackrel{\rightarrow}
    {{\vphantom F}W}_{ia}\!G^{ab}_{\,\,\rm DN}\!
    \stackrel{\leftarrow}
    {{\vphantom F}W}_{bk}\!\big|\big|\;,       \label{6.27}
    \end{eqnarray}
where the double vertical bar denotes the restriction to the brane
with respect to the both points of the two-point kernel, cf. Eq.
(\ref{2.20}) of Sect.2. This restriction, in particular, allows one
to replace $\stackrel{\rightarrow}{W}{\!\vphantom W}^S_{ia}$ by
$\stackrel{\rightarrow}{W}_{ia}$ in view of (\ref{6.11}). As a
result, the quadratic part of the full tree-level action on the
background $g_0$, $g=g_0+\delta g$, takes the form
    \begin{equation}
    \frac12\,\delta^2_g\big(S^{\rm B}[\,G_0(g)\,]+
    S^b(\,g\,)+ S^{\rm b}_{\rm gb}(\,g\,)\big)=
    \frac12\,\delta g^i \,
    \mbox{\boldmath$F$}_{ik}^{\rm DN}\,\delta g^k,    \label{6.28}
    \end{equation}
where $\mbox{\boldmath$F$}_{ik}^{\rm DN}$ is the full brane-to-brane
operator introduced in (\ref{2.19})
    \begin{equation}
    \mbox{\boldmath$F$}_{ik}^{\rm DN}=\,-
    \stackrel{\rightarrow}{W}_{ia}\!G^{ab}_{\,\,\rm DN}\!
    \stackrel{\leftarrow}{W}_{bk}\!\big|\big|
    +S^{\rm b}_{ik}-
    \chi^\mu_i\,c_{\mu\nu}\chi^\nu_k,             \label{6.29}
    \end{equation}
with the gauge-fixed contribution of the brane action (\ref{2.23})
    \begin{equation}
    \kappa_{ik}=S^{\rm b}_{ik}-
    \chi^\mu_i\,c_{\mu\nu}\chi^\nu_k, \,\,\,
    S^{\rm b}_{ik}\equiv
    \frac{\partial^2 S^{\rm b}}
    {\partial g^i\partial g^k},\,\,\,
    \chi^\mu_i\equiv\frac{\partial\chi^\mu}
    {\partial g^i}.            \label{6.30}
    \end{equation}

Finally, substituting (\ref{6.7}) into (\ref{5.2}) and taking the
Gaussian integral over $\delta g^i$ in the vicinity of the
stationary point $g_0$ one finds in the one-loop approximation
    \begin{equation}
    e^{\textstyle i\SGamma(g_0)}=
    \left.
    \frac{\mu_{\rm B}\,{\rm Det}_{\rm D} Q^A_B}
    {\big(\,{\rm Det}_{\rm DN} F_{ab}\,\big)^{1/2}}\,
    \frac{\mu_{\rm b}\,{\rm det}\,
    \mbox{\boldmath$J$}^\mu_\nu}
    {\big(\,{\rm det}\,
    \mbox{\boldmath$F$}_{ik}^{\rm DN}\,\big)^{1/2}}\,
    \exp i \big(\,S^{\rm B}[\,G\,]
    + S^{\rm b}(\,g_0)\,\big)\,
    \right|_{\,G\,=\;G_0(\,g_0)}.              \label{6.32}
    \end{equation}
The preexponential factor here confirms the decomposition property
for the one-loop effective action (\ref{2.17}) advocated in Sect.2
(modulo $\delta(0)$-type terms generated by the local measure
(\ref{4.17})).

\section{Ward identities and gauge independence}
It is important that the both bulk and brane preexponential factors
in (\ref{6.32}) are separately gauge independent. For the bulk
prefactor this is a direct corollary of the gauge independence of
$\mbox{\boldmath$\varPsi$}_{\rm B}(\,g\,)$, (\ref{5.4}). However, it
is worth showing explicitly how this property works in virtue of the
Ward identities for the bulk propagators.

Bulk Ward identities (for tree-level propagators) arise as follows
\cite{Dirac}. Functional differentiation of (\ref{3.3}) shows that
on shell, that is on the background satisfying classical equations
of motion, the functional matrix $S_{ab}$ is degenerate because it
has zero-eigenvalue eigenvectors -- the gauge generators
    \begin{eqnarray}
    R^{\,\,a}_{\!A} S^{\rm B}_{ab}=
    -S^{\rm B}_a\frac{\delta R^a_A}{\delta g^b}=0 \label{7.1}
    \end{eqnarray}
As a consequence the operator $F_{ab}$ satisfies the relation
    \begin{eqnarray}
    R^{\,\,a}_{\!A} F_{ab}=-Q^{\,\,B}_{\!A}
    c_{BC} H^C_{\,\,\,b}               \label{7.2}
    \end{eqnarray}
which can be functionally contracted with matrices of the gauge
$G^{bc}$ and ghost $Q^{-1\,A}_D$ Green's functions. Integration by
parts of the derivatives in the operators
$R^a_A=R^a_{\;\,A}(\partial)$ and $Q^B_A=Q^{\;\,B}_A(\partial)$ does
not produce additional surface terms in view of the boundary
conditions for the propagators. Thus, one arrives at the identity
relating the Dirichlet-Neumann Green's function of the metric
operator to the Dirichlet Green's function of the Faddeev-Popov
operator
    \begin{eqnarray}
    c_{AB}\!\stackrel{\rightarrow}{H}{\!\!\vphantom
    H}^B_{\;\,b}(\partial)\,
    G^{bc}_{\,\rm DN}(y,y')=-Q^{-1\,B}_{\,A}(y,y')\!
    \stackrel{\leftarrow}
    {{\vphantom F}R}
    {\!\!\vphantom R}^{\;\,c}_B(\partial_{y'}).       \label{7.3}
    \end{eqnarray}
It is important that the Dirichlet conditions at $y$ belonging to
the boundary on the right hand side match with the Neumann
conditions (\ref{6.11}) on the left hand side.

Therefore, using the variational definitions of the functional
determinants one has
    \begin{eqnarray}
    &&\delta_H\,{\rm ln}\,{\rm Det}_{\rm DN}\,F_{ab}=
    \delta_H\!\!\stackrel{\leftrightarrow}{{\vphantom F}F}_{ab}\!
    G^{ba}_{\,\rm DN}=
    -2\,c_{AB}\!\stackrel{\rightarrow}{{\vphantom F}H}{\!}^B_b\,
    G^{ba}_{\,\rm DN}\,
    \!\!\stackrel{\leftarrow}
    {\delta H}{\!}^A_a,                         \label{7.4}\\
    &&\delta_H\,{\rm ln}\,
    {\rm Det}_{\rm D}\,Q^A_B
    =Q^{-1\,B}_{\,A}\,\delta_{\!H}\!
    \stackrel{\leftarrow}{{\vphantom F}Q}{\!}^A_B=
    Q^{-1\,B}_{\,A}\!
    \stackrel{\leftarrow}
    {{\vphantom F}R}{\!}^{\,\,\,b}_{\!B}\,
    \!\!\stackrel{\leftarrow}
    {\delta H}{\!}^{\,A}_b,                              \label{7.5}
    \end{eqnarray}
so that in virtue of (\ref{7.3}) the bulk part of the one-loop
effective action is $H$-independent
    \begin{eqnarray}
    \delta_H \!\left(
    -\frac12\,{\rm Tr}_{\rm DN}\ln F
    +{\rm Tr}_{\rm D}\ln Q \right)=0.             \label{7.6}
    \end{eqnarray}

Bulk gauge obviously participates in the construction of the brane
part of the effective action, but the latter turns out to be also
independent of $H^A$. To show this, note first of all that in view
of (\ref{6.11})
    $\stackrel{\rightarrow}{W}_{ia}\!G^{ab}_{\,\,\rm DN}\!
    \stackrel{\leftarrow}{W}_{bk}\!\!||=\,
    \stackrel{\rightarrow}{W}{\vphantom W}^{\!\!\rm S}_{\!\!ia}
    \,G^{ab}_{\,\,\rm DN}\!
    \stackrel{\leftarrow}{W}{\vphantom W}^{\!\!\rm S}_{\!\!bk}||$.
This replacement of the Wronskian operators $W(\partial)$ by their
$H$-independent analogues $W^{\rm S}(\partial)$, cf. Eq.
(\ref{3.25}), implies that
    \begin{equation}
    \delta_H\mbox{\boldmath$F$}_{ik}^{\rm DN}=\,\,
    -\stackrel{\rightarrow}{{\vphantom F}W}{\vphantom W}^{\!\!\rm S}_{\!\!ia}\,
    \big(\delta_H G^{ab}_{\,\,\rm DN}\big)\!
    \stackrel{\leftarrow}{{\vphantom F}W}
    {\vphantom W}^{\!\!\rm S}_{\!\!bk}
    \,\big|\big|\;.                                 \label{7.7}
    \end{equation}
The variation of the Green's function accounting for its boundary
conditions is derived in \ref{Variation}. Using (\ref{C.5}) one has
in view of (\ref{7.3})
    \begin{eqnarray}
    &&\delta_H G^{ab}_{\,\,\rm DN}
    =-G^{ac}_{\,\,\rm DN}\,
    \delta_H \!\!\stackrel{\leftrightarrow}
    {{\vphantom F}F}{\!\!}_{cd}\,
    G^{db}_{\,\,\rm DN}=2G^{(ac}_{\,\,\rm DN}
    \!\stackrel{\leftarrow}{{\vphantom F}H}
    {\!\!\vphantom H}^{\,\;A}_c c_{AB}
    \!\stackrel{\rightarrow}{\delta H}{\!\!\vphantom H}^{B}_{\,\;d}\,
    G^{db)}_{\,\,\rm DN}\nonumber\\
    &&\qquad\qquad\qquad\qquad=
    -2\!\stackrel{\rightarrow}{{\vphantom F}R}
    {\!\vphantom R}^{(a}_{\,A} \,Q^{-1\,A}_{\,\,\,B}
    \!\stackrel{\rightarrow}{\delta H}
    {\!\!\vphantom H}^{B}_{\,\;d}\,
    G^{db)}_{\,\,\rm DN},                                 \label{7.8}
    \end{eqnarray}
so that in virtue of the identity (\ref{3.27}) the argument of the
ghost propagator in
    \begin{eqnarray}
    &&\delta_H\mbox{\boldmath$F$}_{ik}^{\rm DN}=2\!
    \stackrel{\rightarrow}{W}{\vphantom W}^{\!\!\rm S}_{\!\!(ia}\!
    \stackrel{\rightarrow}{R}
    {\!\vphantom R}^{a}_{\,A}
    \,Q^{-1\,A}_{\,\,\,B}
    \delta H^{B}_{\,\;d}\,
    G^{db}_{\,\,\rm DN}\!
    \stackrel{\leftarrow}
    {{\vphantom F}W}_{bk)}\!\big|\big|_{\,i,k}\nonumber\\
    &&\qquad\qquad\qquad\qquad\qquad\qquad
    =-2\,\frac{\partial \mbox{\boldmath$H$}_A}{\partial q^{(i}}
    \,Q^{-1\,A}_{\,\,\,B}
    \delta H^{B}_{\,\;d}\,
    G^{db}_{\,\,\rm DN}\!
    \stackrel{\leftarrow}
    {{\vphantom F}W}_{bk)}\!\Big|\Big|_{\;i,k}=0    \label{7.9}
    \end{eqnarray}
is not differentiated at the boundary because $\partial
\mbox{\boldmath$H$}_A/\partial q^i$ is ultralocal in $y$, and the
whole variation vanishes because of the Dirichlet boundary
conditions for the ghost propagator,
    $(\partial \mbox{\boldmath$H$}_A/\partial q^i)
    \,Q^{-1\,A}_{\,\,\,B}\big|_{\;i}=0$.
Finally,
    \begin{eqnarray}
    \delta_H\!\left(-\frac12\,{\rm tr}\,
    \ln \mbox{\boldmath$F$}^{\rm DN}
    +{\rm tr}\,\ln \mbox{\boldmath$J$}\right)=0.   \label{7.11}
    \end{eqnarray}

Let us turn to the dependence of the effective action on the brane
gauge $\chi^\mu$. As in usual gauge theories, the effective action
is gauge-independent on shell, that is with the sources switched on
the brane off.\footnote{By construction, there are no sources in the
bulk, so that both $\mbox{\boldmath$\varPsi$}_{\rm B}(\,g\,)$ and
$\varGamma$ are universally independent of the bulk gauge $H$,
because the latter is not sensitive to the inclusion of
boundary/brane sources. On the contrary, the brane gauge-fixing
procedure is vulnerable for gauge non-invariant sources on the
brane.} To see this in the one-loop order, note that the tree-level
brane action is identically invariant with respect to
$d$-dimensional (brane) diffeomorphisms
    \begin{equation}
    R^i_\mu\,\frac{\partial}{\partial g^i}\,
    \big(S^{\rm B}[\,G_0(g)\,]+
    S^b(\,g\,)\big)\equiv 0.                           \label{7.12}
    \end{equation}
The differentiation of this identity at $g=g_0$ (mass shell
(\ref{6.16a})) then gives
    \begin{eqnarray}
    &&R^i_\mu\,\Big(-\stackrel{\rightarrow}{{\vphantom F}W}_{ia}\!
    G^{ab}_{\,\,\rm DN}\!
    \stackrel{\leftarrow}{{\vphantom F}W}_{bk}\!\big|\big|
    +S^{\rm b}_{ik}\,\Big)=
    R^i_\mu\,\frac{\partial^2}{\partial g^i\partial g^k}
    \,\Big(S^{\rm B}[\,G_0(g)\,]+
    S^b(\,g\,)\Big)_{g\,=\,g_0}\nonumber\\
    &&\qquad\qquad\qquad\qquad\qquad
    =-\frac{\partial R^i_\mu}{\partial g^k}
    \frac{\partial}{\partial g^i}\,
    \Big(S^{\rm B}[\,G_0(g)\,]+
    S^b(\,g\,)\Big)_{g\,=\,g_0}=0.               \label{7.13}
    \end{eqnarray}
This means that the the brane-to-brane operator (\ref{6.29}) is
non-degenerate entirely due to its brane gauge-breaking term.
Therefore, the following relation holds between the operators
$\mbox{\boldmath$F$}_{ik}^{\rm DN}$ and
$\mbox{\boldmath$J$}^\nu_\mu$
    \begin{equation}
    R^i_\mu\,\mbox{\boldmath$F$}_{ik}^{\rm DN}
    =-\mbox{\boldmath$J$}^\nu_\mu
    \,c_{\nu\alpha}\,\chi^\alpha_k.               \label{7.14}
    \end{equation}
Consequently, their Green's functions satisfy the brane Ward
identity
    \begin{eqnarray}
    &&\mbox{\boldmath$J$}^{-1\,\mu}_{\;\;\nu}\,R^i_\mu
    =-c_{\nu\alpha}\,\chi^\alpha_k\,
    \mbox{\boldmath$G$}^{ki},                       \label{7.15}\\
    &&\mbox{\boldmath$F$}_{ik}^{\rm DN}
    \mbox{\boldmath$G$}^{km}=\delta^m_i,        \label{7.16}
    \end{eqnarray}
where $\mbox{\boldmath$G$}^{ki}$ is the brane-to-brane propagator.
The relation (\ref{7.15}) is a direct analogue of the bulk Ward
identity (\ref{7.3}), except that we don't have to care about
integrations by parts in $x$, because the relevant surface terms are
vanishing at the brane infinity $x^0\to\pm\infty$ (or missing in
Euclidean context in view of the closed nature of the boundary).

Thus, using the variations
    \begin{eqnarray}
    \delta_\chi\mbox{\boldmath$F$}_{ik}^{\rm DN}=
    -2\delta\chi^\nu_{(i}\,c_{\nu\alpha}
    \,\chi^\alpha_{k)},\,\,\,\,
    \delta_\chi \mbox{\boldmath$J$}^\nu_\mu
    =R^i_\mu\,\delta\chi^\nu_i                 \label{7.17}
    \end{eqnarray}
we immediately obtain in view of (\ref{7.15}) the on-shell gauge
independence of the brane effective action
    \begin{eqnarray}
    \delta_\chi\!\left(-\frac12\,{\rm tr}\,
    \ln \mbox{\boldmath$F$}^{\rm DN}
    +{\rm tr}\,\ln \mbox{\boldmath$J$}\right)=
    \delta\chi^\nu_i\,c_{\nu\alpha}\,
    \chi^\alpha_k\,\mbox{\boldmath$G$}^{ki}
    +R^i_\mu\,\delta\chi^\nu_i\,
    \mbox{\boldmath$J$}^{-1\,\mu}_{\;\;\nu}=0.     \label{7.18}
    \end{eqnarray}

\section{Reduction of the Dirichlet-Neumann problem
to the Dirichlet type} If we would apply the stationary phase method
directly to the path integral (\ref{4.14}) without using the bulk
wavefunction representation (\ref{5.2}), then the one-loop
contribution of the metric field would be determined by the
functional determinant subject to Neumann boundary conditions for
all metric components. So the decomposition of the integration
procedure (\ref{5.1}) exercises a half-way reduction to the
Dirichlet problem, because the resulting boundary conditions are of
a mixed type. As it was discussed in Introduction and in
\cite{qeastb}, the construction of the Dirichlet Green's function is
much easier than the Neumann one, so actually no calculational
advantages are gained when this reduction is incomplete. Therefore,
it is worth making a further reduction from the Dirichlet-Neumann
problem to the purely Dirichlet one.

For this purpose introduce the Gaussian path integral over metric
perturbations $\delta G^a\equiv\phi^a=(\phi^i,\phi^A)$ with slightly
more general than in (\ref{6.6}) -- inhomogeneous -- boundary
conditions, $\phi^i|=\varphi^i$,
    \begin{eqnarray}
    &&Z(\varphi^i)=
    \int\limits_{\phi^i|\,=\,\varphi^i}
    D\phi\,\exp{(-S[\,\phi\,])},                  \label{8.1}\\
    &&S[\,\phi\,]\equiv\frac12\,\int dy\,
    \phi^a
    \stackrel{\leftrightarrow}
    {{\vphantom F}F}_{ab}\phi^b.               \label{8.2}
    \end{eqnarray}
Its stationary point $\phi^a_{\rm DN}(\varphi^i)$ is determined by
the Dirichlet-Neumann problem with the inhomogeneous boundary
conditions (\ref{6.21})-(\ref{6.23}) and has the form (\ref{6.24}).
As shown in \ref{DNGauss}, the gaussian integral itself equals
    \begin{eqnarray}
    Z(\varphi^i)=
    \left(\,{\rm Det}\,G_{\rm DN}\right)^{1/2}
    \exp(-S[\,\phi_{\rm DN}(\varphi^i)\,]),            \label{8.3}
    \end{eqnarray}
where the action at the stationary point is
    \begin{eqnarray}
    S[\,\phi_{\rm DN}(\varphi^i)\,]=
    -\frac12\,\varphi^i\left(
    \stackrel{\rightarrow}{{\vphantom F}W}_{ia}\!G_{\rm DN}^{ab}\!
    \stackrel{\leftarrow}{{\vphantom F}W}_{bk}\!||\right)
    \varphi^k\,.                  \label{8.4}
    \end{eqnarray}

Alternatively it can be calculated by integrating first over the
fields subject to Dirichlet boundary conditions for {\em all}
$\phi^a$ with the integration over $\varphi^A\equiv\phi^A|$ reserved
to the last
    \begin{eqnarray}
    Z(\varphi^i)=\int \prod_A\,d\varphi^A
    \int\limits_{\phi^a|\,=\,\varphi^a}
    D\phi\,\exp{(-S[\phi])}.              \label{8.5}
    \end{eqnarray}
The inner integral
    \begin{eqnarray}
    \int\limits_{\phi^a|\,=\,\varphi^a}
    D\phi\,\exp{(-S[\phi])}=
    \left(\,{\rm Det}\,G_{\rm D}\,\right)^{1/2}
    \exp(-S[\,\phi_{\rm D}(\varphi^a)\,])              \label{8.6}
    \end{eqnarray}
is determined by the Dirichlet problem for the field $\phi^a_{\rm
D}(\varphi)$, $F_{ab}\phi^b_{\rm D}=0$, $\phi^a_{\rm
D}\,|=\varphi^a$, having as a solution the following expression in
terms of the Dirichlet Green's function $G_{\rm D}$ of the operator
$F_{ab}$
    \begin{eqnarray}
    \phi^a_{\rm D}(\varphi)=-G^{ab}_{\rm D}\!
    \stackrel{\leftarrow}{W}_{\!bc}\!
    \varphi^c \big|_{\,c}\;.                   \label{8.7}
    \end{eqnarray}
The action in (\ref{8.6}) equals
    \begin{eqnarray}
    &&S[\,\phi_{\rm D}(\varphi^a)\,]=
    \frac12\,\varphi^a\,
    \mbox{\boldmath$D$}_{\,ab}\,
    \varphi^b,                \label{8.8}\\
    &&\mbox{\boldmath$D$}_{\,ab}\equiv\;
    -\stackrel{\rightarrow}
    {{\vphantom F}W}_{ac}\!G_{\rm D}^{cd}
    \stackrel{\leftarrow}{{\vphantom F}W}_{db}
    \!\big|\big|\, ,                      \label{8.8a}
    \end{eqnarray}
so that the gaussian integral over the boundary field $\varphi^A$ in
(\ref{8.5}) is saturated by the saddle point $\varphi^A_0$ of the
above quadratic form in $\varphi^A$,
    $\varphi^A_0=
    -\mbox{\boldmath$G$}^{\,AB}
    \mbox{\boldmath$D$}_{B i} \varphi^i$,
where $\mbox{\boldmath$D$}_{B i}$ is the $B i$-block of the operator
(\ref{8.8a}) and $\mbox{\boldmath$G$}^{\,AB}$ is the inverse of its
$AB$-block
    \begin{eqnarray}
    \mbox{\boldmath$D$}_{AB}\,
    \mbox{\boldmath$G$}^{BC}
    =\delta^C_A                                 \label{8.11}
    \end{eqnarray}
(note that the operator (\ref{8.8a}) differs from the kernel of the
action (\ref{8.4}) by the type of the Green's function -- Dirichlet
vs Dirichlet-Neumann one).

Thus, integration over $\varphi^A$ in (\ref{8.5}) gives
    \begin{eqnarray}
    &&Z(\varphi^i)=\Big(\,{\rm Det}\,G_{\rm D}\Big)^{1/2}
    \Big(\,{\rm det}\,\mbox{\boldmath$D$}_{AB}
    \Big)^{-1/2}
    \exp(-S[\,\phi_{\rm D}(\varphi_0)\,]),      \label{8.12}\\
    &&S[\,\phi_{\rm D}(\varphi_0)\,]=
    \frac12\,\varphi^i\left(
    \mbox{\boldmath$D$}_{ik}-
    \mbox{\boldmath$D$}_{i A}\,
    \mbox{\boldmath$G$}^{\,AB}\,
    \mbox{\boldmath$D$}_{B k}
    \right)\varphi^k.                          \label{8.13}
    \end{eqnarray}
Comparison with (\ref{8.3})-(\ref{8.4}) then gives the following two
relations
    \begin{eqnarray}
    &&-\big( \stackrel{\rightarrow}
    {{\vphantom W}W}\!G_{\rm DN}\!
    \stackrel{\leftarrow}
    {{\vphantom W}W}\!||\,\big)_{\,ik}=
    \mbox{\boldmath$D$}_{ik}-
    \mbox{\boldmath$D$}_{i A}\,
    \mbox{\boldmath$G$}^{AB}
    \mbox{\boldmath$D$}_{B k},                \label{8.14}\\
    &&\Big(\,{\rm Det}\,G_{\rm DN}\Big)^{1/2}
    =\Big(\,{\rm Det}\,G_{\rm D}\Big)^{1/2}
    \Big(\,{\rm det}\,
    \mbox{\boldmath$D$}_{AB}\Big)^{-1/2}.    \label{8.15}
    \end{eqnarray}
They allow one to completely reduce the algorithm (\ref{6.32}) to
that of the Dirichlet boundary conditions. In particular, the
substitution of (\ref{8.14}) into the expression for the
brane-to-brane operator $\mbox{\boldmath$F$}_{ik}^{\rm DN}$,
(\ref{6.29}), expresses its functional determinant as
    \begin{eqnarray}
    &&{\rm det}\,\mbox{\boldmath$F$}_{ik}^{\rm DN}
    ={\rm det}\,\Big((\mbox{\boldmath$D$}_{ik}+\kappa_{ik})
    -\mbox{\boldmath$D$}_{i A}\,
    \mbox{\boldmath$G$}^{AB}
    \mbox{\boldmath$D$}_{B k}\Big)
    ={\rm det}\,\mbox{\boldmath$F$}^{\rm D}_{ab}\;
    \Big(\,{\rm det}\,\mbox{\boldmath$D$}_{\!AB}
    \Big)^{-1},                                    \label{8.16}
    \end{eqnarray}
where $\mbox{\boldmath$F$}^{\rm D}_{ab}$ is the brane-to-brane
operator of another type -- it acts on the space of all metric
perturbations on the brane $\delta G^a|=(\delta g^i,\delta N^A)|$,
$a=(i,A)$, and is built of the Dirichlet Green's function rather
than the mixed-type one
    \begin{eqnarray}
    \mbox{\boldmath$F$}^{\rm D}_{ab}
    =\left[\begin{array}{cc}
    \,\mbox{\boldmath$D$}_{ik}+\kappa_{ik}\,
    &\;\mbox{\boldmath$D$}_{i B}\\
        \mbox{\boldmath$D$}_{A k}
        &\;\mbox{\boldmath$D$}_{AB}
    \end{array}\right]
    =-\big( \stackrel{\rightarrow}
    {{\vphantom W}W}\!G_{\rm D}\!
    \stackrel{\leftarrow}
    {{\vphantom W}W}\!||\,\big)_{\,ab}
    +\kappa_{ik}\,\delta^i_a\delta^k_b.      \label{8.17}
    \end{eqnarray}
Its $ik$-block is given by Eq.(\ref{2.19a}) in Sect.2.

In view of the relation (\ref{8.16}) the one-loop contribution of
the bulk and brane gravitons in (\ref{6.32}) can be decomposed into
the product of other two bulk and brane factors which are entirely
based on the Dirichlet-type objects
    \begin{eqnarray}
    \Big(\,{\rm Det}_{\rm DN}\,F\Big)^{-1/2}
    \Big(\,{\rm det}\,\mbox{\boldmath$F$}_{ik}^{\rm DN}
    \Big)^{-1/2}
    =\Big(\,{\rm Det}_{\rm D}\,F\Big)^{-1/2}
    \Big(\,{\rm det}\,
    \mbox{\boldmath$F$}^{\rm D}_{ik}\Big)^{-1/2}.  \label{8.18}
    \end{eqnarray}
In fact, this decomposition literally repeats the Neumann-Dirichlet
reduction suggested in \cite{qeastb}. The corresponding
decomposition of the one-loop effective action has the form
(\ref{8.19}) presented in Sect.2. From calculation viewpoint it is
simpler than (\ref{2.17}), but it destroys manifest gauge
independence of the combined bulk-brane diagrammatic technique.
Whereas in (\ref{2.17}) both bulk and brane parts are separately
gauge-independent, no such property holds for (\ref{8.19}) -- only
the sum of bulk and brane terms is independent of the bulk gauge
conditions.

\section{Conclusions}
Thus we have an exhaustive formulation of the
Feynman-DeWitt-Faddeev-Popov gauge-fixing procedure in gravitational
systems with branes or boundaries. This procedure incorporates a
special choice of gauge conditions in the bulk and on the brane,
which separately fix the bulk and brane diffeomorphisms. Also it
establishes the boundary conditions for the corresponding ghost
factors and allows one to construct a special bulk wavefunction
representation of the brane effective action
(\ref{5.2})-(\ref{5.3}). The bulk wavefunction satisfies the
generalized Wheeler-DeWitt equations with respect to the induced
metric of the brane (\ref{5.6}). They might, perhaps, serve as a
basis for non-perturbative methods in brane models, alternative to
semiclassical expansion \cite{gospel}.

We derived the boundary conditions for propagators of the Feynman
diagrammatic technique in brane models and explicitly built the
brane effective action in the one-loop approximation. Similarly to
non-gauge (quantized matter) models with branes, considered in
\cite{qeastb}, this one-loop action can be decomposed into the sum
of bulk and brane contributions both in the graviton and the ghost
sectors (\ref{8.19}). The bulk contribution is represented by the
usual functional determinant of the propagator in the
$(d+1)$-dimensional bulk, subject to Dirichlet boundary conditions
on the brane. The brane contribution is given by a similar
determinant of the brane-to-brane operator (\ref{2.19a}) in the
surface $d$-dimensional theory. In the graviton sector this operator
has a nonlocal (pseudodifferential) nature, whereas in the ghost
sector for local gauges this is always a local Faddeev-Popov
operator associated with gauging away the $d$-dimensional surface
diffeomorphisms. This property follows from the generic bulk-brane
factorization of a gauge-fixing factor (\ref{2.5}) in the path
integral for gravitational brane models, which holds beyond loop
expansion.

Linear algebra manipulations and Gaussian integrations which underly
the above results look innocent at the calculational level adopted
in this paper. The obtained algorithms are, however, marred by
ultraviolet divergences and should be regulated by some covariant
technique. The efficiency and correctness of these calculations was,
nevertheless, confirmed within the dimensional regularization in
\cite{qeastb}. In particular, correct expressions for surface
Schwinger-DeWitt coefficients were obtained in \cite{qeastb} and,
thus, guaranteed correct renormalization of logarithmic divergences
in simplest brane models.

Such calculations should also apply in gravitational systems
considered above, and this is a subject of forthcoming papers. We
plan to make a synthesis of the Neumann-Dirichlet reduction with a
systematic curvature expansion method in brane models in order to
alleviate their formalism to the level of universality of the
Schwinger-DeWitt technique \cite{PhysRep,CPTI}. This is important
for various applications in quantum gravity and cosmology, including
quantum consistency of brane models \cite{NicolisRattazzi}, the
conformal anomaly mechanism of the dark energy, that might be
facilitated within the brane concept of extra dimensions
\cite{slih}, and the others.

\appendix
\renewcommand{\thesection}{Appendix \Alph{section}}
\renewcommand{\theequation}{\Alph{section}.\arabic{equation}}

\section{The local measure in brane models\label{Measure}}
Let the bulk and brane terms of the action both have kinetic terms
quadratic in velocities with coefficients $a(X)$ and $b(x)$. These
terms can be rewritten as one integral over the bulk
    \begin{eqnarray}
    &&\int_{\rm\bf B} dX\,a(X)\,(\partial_t\phi(X))^2
    +\int_{\rm\bf b} dx\,b(x)\,
    (\partial_t\varphi(x))^2\nonumber\\
    &&\qquad\qquad\qquad\qquad\qquad
    =\int_{\rm\bf B} dX\,dX'\,{\cal G}(X,X')\,
    \partial_t\phi(X)\,\partial_t\phi(X')
    \end{eqnarray}
with the overall coefficient of the quadratic form in field
velocities
    \begin{eqnarray}
    {\cal G}(X,X')=\delta(x,x')
    \left[\,a(X)\,\delta(y-y')
    +b(x)\,\delta(y)\,\delta(y')\,\right].  \label{A.2}
    \end{eqnarray}
Here $\phi(X)$ and $\varphi(x)=\phi(x,0)=\phi(X)|$ denote the bulk
field and its boundary value at the brane located at $y=0$.

The full local canonical measure reads in terms of the functional
determinant of this matrix as
    \begin{eqnarray}
    \mu[\,a(X),b(x)\,]\equiv\Big({\rm Det}\,
    {\cal G}(X,X')\Big)^{1/2}=
    \mu_B[\,a(X)\,]\,\mu_{\,b}[\,b(x)\,]     \label{A.3}
    \end{eqnarray}
and as we show below factorizes into the product of the
corresponding bulk $\mu_B[\,a(X)\,]$ and brane $\mu_{\,b}[\,b(x)\,]$
measures given by
    \begin{eqnarray}
    \mu_B[\,a(X)\,]=\left(\,\frac{{\rm Det}_{(d+1)}\,
    a(X)\,\delta(X,X')}
    {{\rm Det}_{(d)}\,a(x,0)\,
    \delta(x,x')}\,\right)^{1/2}                    \label{A.4}
    \end{eqnarray}
and
    \begin{eqnarray}
    \mu_{\,b}[\,b(x)\,]=\left(\,{\rm Det}_{(d)}
    \,b(x)\,\delta(x,x')\,\right)^{1/2}            \label{A.5}
    \end{eqnarray}

This factorization follows from the expression for the inverse of
the functional matrix (\ref{A.2})
    \begin{eqnarray}
    {\cal G}^{-1}(X,X')=\delta(x,x')
    \frac1{a(X)}\left[\,\delta(y-y')
    -\frac{b(x)\,\delta(y)\,
    \delta(y')}{a(x,0)+\delta(0)\,b(x)}\,\right].
    \end{eqnarray}
When substituted into the variational equation for the logarithm of
the functional determinant in (\ref{A.3}) this expression yields the
decomposition into the following sum of two terms
    \begin{eqnarray}
    &&\delta \ln {\rm Det}\,{\cal G}(X,X')=\int dX\,dX'\;
    \delta{\cal G}(X,X')\,{\cal G}^{-1}(X',X)\nonumber\\
    &&\qquad=\delta\left[\,\delta^{(d+1)}(0)\int_{\rm\bf B}
    dX\,\ln\,a(X)+\delta^{(d)}(0)\int_{\rm\bf b} dx\,
    \ln\left(1+\delta(0)\frac{b(x)}{a(x,0)}\right)\right].
    \end{eqnarray}
With the obvious limit\footnote{Any regularization implies this
limit except the dimensional regularization in which $\delta(0)=0$,
but in the dimensional regularization the contribution of the local
measure is identically zero.} $\delta(0)\to\infty$ this sum gives
the factorization of the functional determinant
    \begin{eqnarray}
    {\rm Det}\,{\cal G}(X,X')=
    \frac{{\rm Det}_{(d+1)}\,a(X)\,\delta(X,X')}
    {{\rm Det}_{(d)} \,a(x,0)\,\delta(x,x')}\,\,
    {\rm Det}_{(d)} \,b(x)\,\delta(x,x'),
    \end{eqnarray}
which implies the factorization (\ref{A.3})-(\ref{A.5}). In its turn
this confirms the factorization of the local measure in the brane
model (\ref{4.17})-(\ref{4.19}). The division by ${\rm
Det}_{(d)}\,a(x,0)\,\delta(x,x')$ in the bulk measure (\ref{A.4})
corresponds to the absence of local factors belonging to the brane
in the product of (\ref{4.18}). In the notations of this Appendix
this means the following chain of relations
    \begin{eqnarray}
    &&\prod\limits_{X,\,X\not\in {\rm\bf b}} a(X)
    =\prod\limits_{y>0,\,x} a(x,y)
    =\prod\limits_x \frac1{a(x,0)}\,
    \prod\limits_{y\geq 0}\, a(x,y)\nonumber\\
    &&\qquad=
    \prod\limits_x \frac1{a(x,0)}\,\exp\left(\delta(0)
    \int_{y\geq 0} dy\,\ln a(x,y)\right)=
    \frac{{\rm Det}_{(d+1)}\,a(X)\,\delta(X,X')}
    {{\rm Det}_{(d)} \,a(x,0)\,\delta(x,x')}\,.
    \end{eqnarray}
Note that this division is responsible for zero density weight of
the bulk wavefunction $\mbox{\boldmath$\varPsi$}_{\!\rm B}(g)$,
defined by the path integral (\ref{5.3}) (see footnote
\ref{footnote} in Sect.5).

\section{The Gaussian functional integral with mixed
boundary conditions \label{DNGauss}} Feynman's calculation
\cite{Feynman} of the gaussian functional integral with the action
(\ref{6.6}) for the case when only a part of integration fields is
fixed at the boundary,
    \begin{eqnarray}
    \delta G^a\equiv \phi^a
    =(\phi^i,\,\phi^A),\,\,\,\,
    \phi^i\big|=0,                     \label{B.0}
    \end{eqnarray}
can be based on the integral
    \begin{eqnarray}
    &&Z[\,F,J\,]=
    \int\limits_{\phi^i |\,=\,0}
    D\phi\,\exp{(-S[\,\phi,J\,])},              \label{B.1}\\
    &&S[\,\phi,J\,]=\int dy\,
    \left(\,\frac12\,\phi^a(y)\!
    \stackrel{\,\leftrightarrow}{{\vphantom F}F}_{ab}
    \!(\partial)\,\phi^b(y)
    +J_a(y)\,\phi^a(y)\right),               \label{B.2}
    \end{eqnarray}
where the action is modified by the source term in the bulk. To find
the dependence of (\ref{B.1}) on $J_a$ consider the stationary point
$\phi_{\rm DN}=\phi_{\rm DN}[J\,]$ of this action with respect to
variations of $\phi^a$ satisfying (\ref{B.0}).
    \begin{eqnarray}
    \delta S[\,\phi,J\,]=\int dy\;\delta\phi^a\,
    (\,\stackrel{{\vphantom F}
    \rightarrow}{{\vphantom F}F}\!\phi+J\,)_a+
    \left.\delta\phi^A\,
    (\stackrel{\rightarrow}
    {{\vphantom W}W}\!\phi)_A \right|=0     \label{B.3}
    \end{eqnarray}
Thus, $\phi_{\rm DN}$ satisfies the problem with mixed
Dirichlet-Neumann boundary conditions
    \begin{eqnarray}
    &&\stackrel{\,\rightarrow}{{\vphantom F}F}_{ab}
    \!\phi^b_{\rm DN}+J_a=0,  \\
    &&\phi^i_{\rm DN} \big|=0,\,\,\,\,\,
    \stackrel{\rightarrow}{{\vphantom W}W}_{\!A a}
    \phi^a_{\rm DN}\big|=0,                       \label{B.3a}
    \end{eqnarray}
and therefore can be represented in terms of the corresponding
Green's function of Eqs.(\ref{6.9})-(\ref{6.11})
    \begin{eqnarray}
    \phi_{\rm DN}^a[J\,]=G^{ab}_{\rm DN} J_{\,b}\,
    \equiv\,\int dy'\,
    G^{ab}_{\rm DN}(y,y')\,J_b(y')             \label{B.3b}
    \end{eqnarray}

Now make the shift of the integration variable in (\ref{B.1}) by
$\phi_{\rm DN}$, $\phi=\phi_{\rm DN}+\Delta$. Under this replacement
the action decomposes in the part $S[\,\Delta,0\,]$ quadratic in
$\Delta$ and the part independent of $\Delta$. Linear in $\Delta$
term is absent (both in the bulk and on the boundary) in view of the
stationarity of the action at $\phi_{\rm DN}$, so that
    \begin{eqnarray}
    &&S[\,\phi,J\,]=S[\,\Delta,0\,]
    +S[\,\phi_{\rm DN},J\,],                  \label{B.4}\\
    &&S[\,\phi_{\rm DN},J\,]=
    \frac12\,J_{a}\,G^{ab}_{\rm DN} J_{\,b}.        \label{B.4a}
    \end{eqnarray}
Therefore
    \begin{eqnarray}
    Z[\,F,J\,]=Z[\,F,0\,]\,\exp{(-S[\,\phi_{\rm DN},J\,])}.  \label{B.5}
    \end{eqnarray}

To find the prefactor, consider the variation of the integral
(\ref{B.1}) at $J=0$ with respect to the operator $F_{ab}$ and make
the following set of obvious identical transformations using the
above equations (\ref{B.4a}) and (\ref{B.5})
    \begin{eqnarray}
    &&\delta_F Z[\,F,0\,]=-\int D\phi\,
    \left(\,\frac12\,\phi^a\!
    \stackrel{\leftrightarrow}{\delta F}_{ab}\!
    \phi^b \right)
    \exp{(-S[\,\phi,0\,])}\nonumber\\
    &&\nonumber\\
    &&\qquad\qquad\quad=-\left.\int D\phi\,
    \left(\,\frac12\,\frac\delta{\delta J_a}\!
    \stackrel{\leftrightarrow}{\delta F}_{ab}\!
    \frac\delta{\delta J_b}
    \right)\,Z[\,F,J\,]\,\right|_{\,J=0}\nonumber\\
    &&\nonumber\\
    &&\qquad\qquad\quad=
    -\frac12\,\stackrel{\leftrightarrow}{\delta F}_{ab}\!
    G_{\rm DN}^{ba}\,Z[\,F,0\,].              \label{B.6}
    \end{eqnarray}
Here $\stackrel{\leftrightarrow}{\delta F}_{ab}$ means arbitrary
variations of the coefficients of the operator, and the double arrow
implies symmetric action of two first-order derivatives of
$F_{ab}(\partial)$ on both arguments of $G_{\rm DN}^{ba}$ similar to
Eq.(\ref{B.2}).

Therefore, one gets
    \begin{eqnarray}
    &&\delta_F \ln\,Z[\,F,0\,]=
    -\frac12\,\stackrel{\leftrightarrow}{\delta F}_{ab}\!
    G_{\rm DN}^{ba} =-\frac12\,\delta\ln{\rm Det}_{\rm DN} F\equiv
    \frac12\,\delta\ln{\rm Det}\,G_{\rm DN}\,.           \label{A.8}
    \end{eqnarray}
This justifies the variational definition (\ref{6.8}) of the
Dirichlet-Neumann functional determinant of $F_{ab}$ originating
from the Gaussian integration with mixed boundary conditions.

\section{The variation of $G_{\rm DN}^{ab}$ \label{Variation}}
In view of (\ref{6.20}) $\stackrel{\rightarrow}{H}{\!\vphantom
H}^A_a=a^{-1\,AC}\stackrel{\rightarrow}{W}_{Cb}$, and the boundary
conditions in the problem (\ref{6.9})-(\ref{6.11}) for $G_{\rm
DN}^{ab}$ can be rewritten in terms of the Wronskian operator
$\stackrel{\rightarrow}{W}_{Cb}$. Then the variation of this problem
gives the set of inhomogeneous equations for $\delta G_{\rm
DN}^{ab}$
    \begin{eqnarray}
      &&\stackrel{\rightarrow}{F}_{ab} \delta G_{\rm DN}^{bc}
      =-\stackrel{\rightarrow}{\delta F}_{ab} G_{\rm DN}^{bc},\nonumber\\
      &&\stackrel{\rightarrow}{W}_{Ab} \delta G_{\rm
      DN}^{bc}\big|_{\,A}
      =-\stackrel{\rightarrow}{\delta W}_{\!Ab}
      G_{\rm DN}^{bc}\big|_{\,A},\nonumber\\
      &&\delta G_{\rm DN}^{ic}\big|_{\, i}=0         \label{C.1}
      \end{eqnarray}
This set of equations forms the boundary value problem, analogous to
(\ref{6.21})-(\ref{6.24}), but with the inhomogeneous subset of
Neumann boundary conditions
      \begin{eqnarray}
      &&\stackrel{\rightarrow}{F}\!\phi
      =J,\nonumber\\
      &&\big(\!\stackrel{\rightarrow}{W}\!\phi\big)_A\big|
      =w_A,\,\,\,\,\,\phi^i\big|=0.         \label{C.2}
      \end{eqnarray}
It has a solution
    \begin{eqnarray}
    &&\phi^a(X)=\int_{\rm\bf B} dX'\,G^{ab}_{\,\rm DN}(X,X')\,J_b(X')
    +\int_{\rm\bf b} dx\,G_{\,\,\rm DN}^{aA}(X,e(x))\,w_A(x)\nonumber\\
    &&\qquad\qquad\qquad\qquad\qquad\qquad\qquad
    \equiv G^{ab}_{\,\,\rm DN}\,J_b
    +G^{aA}_{\,\,\rm DN}\,w_A \Big|_{\,A}         \label{C.3}
      \end{eqnarray}
which, when applied to (\ref{C.1}), gives
    \begin{eqnarray}
    \delta G_{\rm DN}^{ab}=-G^{ac}_{\,\,\rm DN}\!
    \stackrel{\rightarrow}{\delta F}_{\!\!cd}\!G_{\rm DN}^{db}
    -G^{aA}_{\,\,\rm DN}
    \stackrel{\rightarrow}{\delta W}_{\!\!Ad}\!
      G_{\rm DN}^{db}\Big|_{\,A}
      \end{eqnarray}
Therefore, in view of the relation (\ref{6.10a}), the variation of
the Dirichlet-Neumann Green's function is given by the equation
    \begin{eqnarray}
    \delta G_{\rm DN}^{ab}=-G^{ac}_{\,\,\rm DN}\!
    \stackrel{\leftrightarrow}{\delta F}_{\!\!cd}
    \!G_{\rm DN}^{db},                                    \label{C.5}
    \end{eqnarray}
used in Eq.(\ref{7.8}).

\section*{Acknowledgements}
The author is grateful for hospitality of the Physics Department of
the University of Bologna and the Theoretical Physics Institute of
the University of Cologne. This work was supported by the Russian
Foundation for Basic Research under the grant No 05-01-00996 and the
grant LSS-4401.2006.2.


\begin{thebibliography}{99}
\bibitem{qeastb}A.O.Barvinsky and D.V.Nesterov,  Phys. Rev. {\bf D73} (2006)
066012, hep-th/0512291.
\bibitem{GarrigaTanaka}J.Garriga and T.Tanaka
Phys. Rev. Lett. {\bf 84} (2000) 2778, hep-th/9911055.
\bibitem{ghosts}L.Pilo, R.Rattazzi and A.Zaffaroni, JHEP {\bf
0311} (2000) 056, hep-th/0004028.
\bibitem{ghoststachions}S.L.Dubovsky and V.A.Rubakov, Phys. Rev. {\bf D 67}
(2003) 104014, hep-th/0212222; S.L.Dubovsky and M.V.Libanov, JHEP
{\bf 0311} (2003) 038, hep-th/0309131.
\bibitem{scale}M.A.Luty, M.Porrati and R.Ratazzi, JHEP {\bf 0309} (2003)
029, hep-th/0303116.
\bibitem{NicolisRattazzi}A.Nicolis and R.Rattazzi, JHEP {\bf 06}
(2004) 059, hep-th/0404159.
\bibitem{slih}A.O.Barvinsky and A.Yu.Kamenshchik, {\em Cosmological
landscape from nothing: some like it hot}, hep-th/0605132.
\bibitem{RS}L.Randall and S.Sundrum, Phys. Rev. Lett.
{\bf 83} (1999) 4690, hep-th/9906064.
\bibitem{DGP}G.R.Dvali, G.Gabadadze and M.Porrati, Phys. Lett.
{\bf B485} (2000) 208, hep-th/0005016.
\bibitem{Tanaka}T.Tanaka, Phys. Rev. {\bf D69} (2004) 024001,
gr-qc/0305031.
\bibitem{other}M.Kolanovic, M.Porrati and J.W.Rombouts, Phys.
Rev. {\bf D68} (2003) 064018; M.Porrati and J.W.Rombouts, Phys. Rev.
{\bf D69} (2004) 122003, hep-th/0401211.
\bibitem{DW}B.S.DeWitt, {\em Dynamical Theory of Groups and
Fields} (Gordon and Breach, New York, 1965); Phys.Rev. {\bf 162}
(1967) 1195.
\bibitem{duality}A.O.Barvinsky and D.V.Nesterov, Nucl. Phys.
{\bf B654} (2003) 225, hep-th/0210005.
\bibitem{gospel}A.O.Barvinsky, {\em The Gospel according to DeWitt
revisited: quantum effective action in braneworld models},
hep-th/0504205.
\bibitem{FV}E.S.Fradkin and G.A.Vilkovisky, Phys.Lett. {\bf B55}
(1975) 224, CERN Report TH-2332 (1977).
\bibitem{Leutw}H.Leutwyler, Phys. Rev. {\bf 134} (1964) B1155.
\bibitem{HH}J.B.Hartle and S.W.Hawking, Phys.Rev. {\bf D28} (1983)
2960.
\bibitem{barvin}A.O.Barvinsky and V.N.Ponomariov, Phys.Lett. {\bf B167}
(1986) 289; A.O.Bar\-vin\-sky, Phys.Lett. {\bf B195} (1987) 344;
Phys. Rep. {\bf 230} (1993) 237.
\bibitem{HalHar}J.J.Halliwell and J.B.Hartle, Phys.Rev. {\bf D43} (1991)
1170.
\bibitem{Dirac}A.O.Barvinsky, Nucl. Phys. {\bf B 520} (1998) 533.
\bibitem{operator}A.O.Barvinsky, Phys. Lett. {\bf B241} (1990) 201;
{\em Geometry of the Dirac and reduced phase space quantization of
constrained systems}, gr-qc/9612003; A.O.Barvinsky and V.Krykhtin,
Class.Quantum Grav. {\bf 10} (1993) 1957.
\bibitem{Feynman}R.P.Feynman, Phys.Rev. {\bf 84} (1951) 108.
\bibitem{PhysRep}A.O. Barvinsky and G.A. Vilkovisky,
Phys. Rep. 119 (1985) 1.
\bibitem{CPTI}A.O.Barvinsky and G.A.Vilkovisky, Nucl. Phys.
{\bf B 282} (1987) 163, Nucl. Phys. {\bf B333} (1990) 471;
A.O.Barvinsky, Yu.V.Gusev, G.A.Vilkovisky and V.V.Zhyt\-nikov, J.
Math. Phys. {\bf 35} (1994) 3525; J. Math. Phys. {\bf 35} (1994)
3543.
\end{thebibliography}
\end{document}